\documentclass[aps,pra,showpacs,floatfix,superscriptaddress]{revtex4}
\usepackage{graphicx}
\usepackage{color}
\usepackage{subfigure}
\usepackage{ulem}
\usepackage{epsfig}
\usepackage{comment}
\begin{document}
\title{A thermodynamic geometric study of R\'{e}nyi and Tsallis entropies}
\author{B. N. Tiwari \footnote{BNT acknowledges the local hospitality of the
``\textit{Department of Physics, Indian Institute of Technology
Kanpur, India}".}} \email{bntiwari.iitk@gmail.com}
\affiliation{\small INFN-Laboratori Nazionali di Frascati, Via E.
Fermi 40, 00044 Frascati, Italy}
\author{Vinod Chandra}
\email{vinodc@theory.tifr.res.in}
\affiliation{\small Department of Theoretical Physics,
Tata Institute of Fundamental Rresearch, Homi Bhabha Road, Mumbai-400005, India}
\author{Subhashish Banerjee}
\email{subhasish@cmi.ac.in} \affiliation{\small Chennai
Mathematical Institute, Padur PO, Siruseri- 603103, India}
\affiliation{\small Indian Institute of Technology Rajasthan, Jodhpur-342011, India}

\begin{abstract}
 A general investigation is made into the
 intrinsic  Riemannian geometry for complex systems, from the perspective of statistical
  mechanics.   The
  entropic formulation of statistical mechanics is the ingredient which
  enables a connection between statistical mechanics and the corresponding
  Riemannian geometry. The form of the entropy used commonly is the Shannon
  entropy. However, for modelling complex systems, it is often useful to make
  use of   entropies such  as the  R\'{e}nyi  and   Tsallis  entropies.
   We  consider, here,   Shannon, R\'{e}nyi, Tsallis, Abe and structural
  entropies, for our  analysis. We focus on   one, two and  three
  particle thermally excited configurations. We find that statistical pair
  correlation functions, determined by the components of the covariant metric tensor
  of the underlying thermodynamic geometry, associated with the various entropies have well defined, definite expressions,
  which may be extended for  arbitrary finite  particle systems. In all
  cases, we find a non-degenerate intrinsic Riemannian manifold. In particular,
  any finite  particle system described in terms of R\'{e}nyi, Tsallis, Abe and structural entropies,
  always corresponds to an interacting statistical system, thereby highlighting their importance in the study of
  complex systems. On the other hand,  a statistical
  description by the Gibbs-Shannon entropy corresponds to a non-interacting system.
\end{abstract}
\maketitle
\section{ Introduction}

Entropy is one of the cornerstones of statistical mechanics,  and is used extensively in many studies.
Apart from the usual Gibbs-Shannon  entropy, there is a growing consensus that for an understanding of
complex systems, it would be useful to go beyond the Shannon entropy. For this reason, the Tsallis \cite{ct88}
and R\'{e}nyi entropies \cite{cb09, bash}, have been used extensively.
On the other hand,  an elegant geometrical formulation of thermodynamics has been
developed, see \cite{rup} for a review. In this, the theory of thermodynamic fluctuations is developed
from a macroscopic perspective,  making use of the notions of covariance and consistency,  expressed
naturally using the language of Riemannian geometry.

Motivated by a need to understand complex systems,  we make a
study of  the intrinsic Riemannian geometry associated with the
usual Gibbs-Shannon as well as the R\'{e}nyi and Tsallis
entropies. In addition, we also study the Abe \cite{sa97} and the
so called structural \cite{kz03} entropies.  This provides a
geometric insight into the systems, studied using these entropies.
Thus, for example, the components of the covariant metric tensor,
of the underlying Riemannian geometry of the thermodynamic phase
space,  provides insight into the local correlations inherent in
the system, which could be used to study its local stability.  On
the other hand, the corresponding thermodynamic scalar curvature
is a signature of the global  correlations present, and could be
used to address issues related to phase transitions, in the
system.

The plan of the paper is as follows.  In Section II, we make a brief review of thermodynamic geometry.  Thermodynamic
geometry is applied  in Section III,  to a study of various entropies, for  few particle systems. The entropies studied are
the Shannon entropy, as well as the R\'{e}nyi, Tsallis, Abe and structural entropies.  This is followed up in Section IV,
by a geometrical interpretation of the additivity of R\'{e}nyi and pseudo-additivity of Tsallis entropies. Finally, in Section V,
we make our Conclusions.

\section{Thermodynamic geometry: A brief Review}

This section provides a brief account of Ruppeiner geometry, which
as an intrinsic Riemannian manifold $(M, g)$, serves the purpose
of describing the nature of statistical fluctuations in an open
thermodynamic configuration. Before focussing our attention on a
specific configuration, it is worthwhile reviewing some of the
basics of intrinsic Riemannian geometry.
More details can be found in
\cite{Weinhold1,Weinhold2,rup,RuppeinerPRD78,gr-qc/0512035v1,gr-qc/0601119v1,
gr-qc/0304015v1,0510139v3,0606084v1} for black holes in general
relativity, \cite{BNTBull,bnt, SST, BNTBull08} for a recent
account of these notions in string theory and M-theory and
\cite{BNTSBVC} for associated chemical correlations and quark
number susceptibilities in $2$- and $3$-flavor hot QCD
configurations.

In what follows,  we shall consider an intrinsic Riemannian geometric model
whose covariant metric tensor may be defined as the Hessian matrix of the entropy
with respect to an arbitrary number of parameters characterizing the thermodynamics of the system
of interest.
We shall focus our attention on  the parameters, such as fixed volume,
characterizing the thermodynamics of  the equilibrium configuration.
In particular, let us define a representation $ S(q,p_i) $ for a given entropy,
generalized temperature, and probabilities $ \lbrace S, q, p_i \rbrace $. Here
$p_i$ is the probability that a given event $i$ has occurred, and
would be determined, for an equilibrium distribution, by its temperature.
These $p_i$'s, for $i = 1,...,n$, when treated as a set of extensive thermodynamic
variables, form coordinate charts for the corresponding intrinsic manifold.
Also, an appropriate choice of the parameter $q$, characterizes the entropy
of the system.


The probability distribution of thermodynamic fluctuations,
in an equilibrium intrinsic space, naturally characterizes the invariant
interval of the corresponding thermodynamic geometry. It can be shown
that the probability distribution in the Gaussian approximation is
\begin{eqnarray}
\Omega(x)= A \ \exp \bigg(-\frac {1}{2}g_{ij}(x) dx^i \otimes dx^j
\bigg),
\end{eqnarray}
where the pre-factor $A$ is the normalization constant and $
\otimes$ signifies symmetric product of the Gaussian distribution.
The associated inverse metric may thus be shown to be the second
moment of the quadratic fluctuations or the pair correlation
functions. An explicit evaluation shows the components of the
inverse metric tensor to be
\begin{eqnarray}
g^{ij}=  \langle X^i \vert X^j\rangle,
\end{eqnarray}
where  $\lbrace X_i \rbrace $'s are the extensive thermodynamic variables conjugate
to the intensive variables   $\lbrace x_i \rbrace $. Moreover, such Riemannian structures
may as well be expressed in terms of any suitable thermodynamic potential obtained by
Legendre transform(s), which in our intrinsic geometric set-up corresponds to certain
general coordinate transformations on the equilibrium statistical configuration.

We shall apply the above formalism, in this work, to a  fixed number of particles.
This method, in general, implies that
the quadratic fluctuations in an underlying statistical system may be described by the
formalism of intrinsic Riemannian geometry, which is solely based on nothing other
than the relative probabilities and generalized temperature. This, in turn,  implies that
the space spanned by the $n$ parameters of the system under consideration exhibits
a n-dimensional intrinsic Riemannian manifold $ (M_n, g)$, the so called thermodynamic
geometry \cite{rup}. The components of the covariant metric tensor may be defined as
\begin{eqnarray} \label{eq1}
 g_{ij}:= -\frac{\partial^2 S(\vec{X})}{\partial X^i \partial X^j},
\end{eqnarray}
where the vector $\vec{X} =(q,p^i) \in M_n $.
Explicitly, for the case of two dimensional intrinsic geometry
parametrized by $\vec{X} =(p^1,p^2) \in M_2 $, the
components of the thermodynamic metric tensor are given by \\
\begin{eqnarray}
\label{eq2}
 g_{p_i p_j}=-\frac{\partial^2 S}{{\partial p_i}{\partial p_j}}\equiv S_{p_i p_j},
\end{eqnarray}
where, $i,j=1,2$. In this case the determinant of the metric tensor may be expressed as,
\begin{eqnarray} \label{determinant}
\Vert g \Vert &= &S_{p_1 p_1}S_{p_2 p_2}- S_{p_1 p_2}^2.
\end{eqnarray}
The Christoffel connection  $\Gamma_{ijk}$,  Riemann curvature tensor $R_{ijkl}$,
Ricci tensor $R_{ij}$
and scalar curvature $ R $ for the two dimensional thermodynamic geometry $(M_2,g)$
can be computed.
The scalar curvature can be shown to be
\begin{eqnarray} \label{curvature}
R&=& \frac{1}{2} \bigg(S_{p_1p_1}S_{p_2 p_2}- S_{p_1 p_2}^2
\bigg)^{-2} \bigg(S_{p_2 p_2}S_{p_1p_1p_1}S_{p_1 p_2 p_2}\nonumber\\
&&+ S_{p_1 p_2}S_{p_1p_1p_2}S_{p_1p_2p_2}+
S_{p_1p_1}S_{p_1p_1p_2}S_{p_2p_2p_2}\nonumber\\ &&-
S_{p_1p_2}S_{p_1p_1p_1}S_{p_2p_2p_2}-
S_{p_1p_1}S_{p_1p_2p_2}^2\nonumber\\&& - S_{p_2p_2}S_{p_1p_1p_2}^2
\bigg).
\end{eqnarray}
Interestingly, the relation between the thermodynamic scalar
curvature and the Riemann curvature tensor for the two dimensional
intrinsic geometry is given (see for details \cite{0606084v1,
bnt}) by
\begin{eqnarray} \label{eq4}
 R=\frac{2}{\Vert g \Vert}R_{p_1p_2p_1p_2}.
\end{eqnarray}
One may further see that this relation is quite natural for any intrinsic Riemannian
surface $(M_2(R),g)$.  Multi-parameter statistical configurations can also be studied
along similar lines.
Physically, it is intriguing
to note that the Riemannian structure defined by the metric tensor, in the entropy representation,
is rather closely related to classical thermodynamic fluctuation theory \cite {Landsberg} and critical
phenomena, if any.  Even though the above geometric formulation thus tacitly involves a microscopic basis in terms
of a chosen ensemble, nevertheless, the present analysis has only been considered in the thermodynamic
limit of open statistical systems.

It is worth noting that the relation of a non-zero scalar
curvature with an underlying interacting statistical system
remains valid for higher dimensional intrinsic manifolds and the
connection of a divergent (scalar) curvature with phase
transitions may as well be analyzed from the Hessian matrix of the
considered fluctuating entropy. Our analysis thus takes into
account  scales that are larger than the correlation length where
a  few microstates cannot dominate the entire macroscopic
phase-space configuration.

We shall focus on the interpretation that the underlying
entropy includes contributions from a large number of microstates and
make use of our description of the geometric thermodynamics of open systems.
With this general introduction to the thermodynamic geometry defined
as the Hessian function of the configurational entropy, we shall now
proceed to investigate the thermo-geometric structures of finite particle
systems, with few parameters. We shall compute entropies
and  study the corresponding thermodynamic geometry.


Having provided a brief description of thermodynamic geometry,
which gives clear prescription for obtaining the metric tensor $\Vert g \Vert $
and associated Ricci scalar curvature $R$, it is worth mentioning that
these quantities find relevance in diverse studies, since the scalar
curvature represents the correlation length of the chosen statistical
configuration \cite{rup,RuppeinerPRD78}. This would be important, as  we shall show
below,  in the context of  complex statistical systems which are seen to be generically interacting.

\section{Thermodynamic geometrical approach to complex systems}

What is presented here may be motivated by saying that a system can be described
by, say, the two states of a coin, \textit{viz.}, heads $H$ or tail $T$.  This would be called, in our nomenclature,
a two particle system,  the classical analog  of a two-level quantum system (qubit),
having two probabilities, taking values over $\{ H,T\}$. Similarly,  a three particle
system can be described by three probabilities, and would be the classical
analog  of a three-level system.
For an entropic description of complex systems, we
shall use the Shannon entropy, as a comparison with,  among others, Tsallis and R\'{e}nyi entropies  \cite{ct88,bash,cb09}.
We shall consider equilibrium thermodynamic systems, and describe their
nature under  Gaussian fluctuations. It will be shown that the
symmetric nature of fluctuations define an intrinsic Riemannian geometry,
whose metric tensor may be obtained as the negative Hessian of the associated
entropy. The number of concerned parameters of a chosen configuration
basically determines the dimensionality of phase-space.

\subsection{Shannon Entropy}
Here we present a brief account of the essential
features of thermodynamic geometries from the perspective of
multi-parameter Shannon configurations. In particular, we shall
focus our attention on the geometric nature of the underlying ensemble
having finitely many particles.
The expression for
Shannon entropy, for arbitrary number of probabilities, is given by
\begin{equation}
S = -\sum\limits_i p_i \ln p_i. \label{shannon}
\end{equation}
Below, we shall compute the concerned configurations for an
arbitrary $i=2,3,...$ and then analyze the generic nature of
statistical correlations in them.
The number of probabilities define
the dimension of the underlying thermodynamic configuration.

\subsubsection{Two particle Shannon system}
Let us first consider the simplest case of Shannon
configuration with two independent probabilities $\{ p_1, p_2 \}$. In this
case, the entropy may  be expressed as
\begin{equation}
S(p_1, p_2):= -p_1 \ln( p_1 ) - p_2 \ln( p_2 ).
\end{equation}
The components of the covariant metric tensor are
explicitly given by
\begin{eqnarray}
g_{ij}=\frac{1}{p_i}\delta_{ij},\ i,j=1,2
\end{eqnarray}
For the two particle Shannon configuration, we find that the determinant
of the metric tensor is non-zero and varies as inverse of the either  of the probabilities.
In particular, we finds that the determinant of the metric tensor is
\begin{equation}
\Vert g \Vert = \frac{1}{ p_1 p_2}.
\end{equation}
Furthermore, it is not difficult to see that the scalar curvature vanishes,
$ R =0 $, and thus the underlying statistical basis is a non-interacting configuration.

\subsubsection{Three particle Shannon system}
An addition of third probability as an intrinsic coordinate to the foregoing
configuration shows that the expression of entropy in this case is
\begin{equation}
S( p_1, p_2, p_3 ):= -p_1 ln( p_1 ) - p_2 ln( p_2 ) - p_3 ln( p_3 ) .
\end{equation}
We may easily observe that the components of the covariant metric tensor are
\begin{eqnarray}
g_{ij}=\frac{1}{p_i}\delta_{ij}; \ i,j=1-3.
\end{eqnarray}
A similar analysis finds non-zero determinant of the metric tensor,
which  takes the form
\begin{equation}
\Vert g \Vert = \frac{1}{ p_1 p_2 p_3}.
\end{equation}
In this case, again, we find that the scalar curvature identically vanishes,
with $R= 0$. Thus, we find that the underlying three particle Shannon
configurations are well-defined and imply non-interacting statistical system.

\subsubsection{ Multi-particle Shannon system}

We further notice  the same conclusions for any finitely many particle Shannon
statistical system.  In  general, we see for multi-particle Shannon systems that
the Shannon pair correlation  functions are
\begin{eqnarray}
g_{ij}=\frac{1}{p_i}\delta_{ij}; \ i,j=1-n.
\end{eqnarray}
It is worth  mentioning that the determinant of the metric tensor is
\begin{equation}
\Vert g \Vert=  \bigg(\prod p_i \bigg)^{-1}.
\end{equation}
An easy inspection finds, for all $i= 2,3,..$, that the corresponding
statistical configurations have zero scalar curvature and are non-interacting.
Our analysis thus shows that the general Shannon configurations are well-defined
thermodynamic systems and have an identically vanishing correlation lengths.

\subsection{R\'{e}nyi Entropy}

In this subsection, we shall use the essential features of
thermodynamic geometry
to describe systems using R\'{e}nyi entropy, with an
increasing number of parameters.
Here, we focus our attention on the geometric nature of the local
and global correlations in the neighbourhood of small fluctuations
in the chosen R\'{e}nyi configurations. As stated earlier, the thermodynamic
metric in the probability space is given by the negative Hessian matrix
of the entropy with respect to the  variables defining the space, which here,
would be  the two, three,... distinct probabilities carried
by the R\'{e}nyi configurations. The entropy
of $n$-particle R\'{e}nyi system \cite{bash} is given by
\begin{equation} \label{RenyiEntropy}
S_q^R =\frac{1}{ 1-q} \ln \bigg(\sum\limits_i p_i^q \bigg).
\label{renyi}
\end{equation}
In order to analyze the underlying interactions, we shall proceed
by considering, first,  the nature of single particle system at given
finite temperature.

\subsubsection{Single particle R\'{e}nyi system}
We shall first consider a single particle R\'{e}nyi system at
temperature $T$. From the Eq. (\ref{RenyiEntropy}), we then see
that the corresponding entropy reduces to
\begin{equation}
 S_q^R(q, p_1 ):= \frac{1}{ 1-q} \ln( p_1^q ), \label{ren}
\end{equation}
Following Eq. (\ref{eq2}), a straightforward calculation Eq. shows
that the components of the covariant metric tensor are
\begin{eqnarray}
g_{qq} &=& 2 \bigg(\frac{\ln( p_1 ) q -\ln( p_1^q ) - \ln( p_1 )}{ (1- q )^3}\bigg), \nonumber\\
g_{qp_1} &=& - \frac{1}{(1- q )^2 p_1}, \nonumber\\
g_{p_1p_1} &=& \frac{1}{(1- q ) p_1^2}.
\end{eqnarray}
Applying  Eq. (\ref{determinant}),
the determinant of the metric tensor can be shown to have the  simple form
\begin{equation}
\Vert g \Vert= \frac{2 \ln( p_1 ) q^2 -2 q \ln( p_1^q ) - 2 \ln(
p_1 ) q - 1}{ (1- q )^4 p_1^2}.
 \end{equation}
Finally, it is not difficult to see from Eq. (\ref{curvature}),
that the Ricci scalar is
\begin{equation}
R=\frac{( -1 + q ) ( \ln( p_1 ) + \ln( p_1^q ) - 2 \ln( p_1 ) q -
q \ln( p_1^q ) + \ln( p_1 ) q^2 + 1 - 2 q )} {(2 q \ln( p_1^q )+ 2
\ln( p_1 ) q- 2 \ln( p_1 ) q^2+ 1 )^2}.
\end{equation}

\begin{figure}
\includegraphics[width=10.0cm]{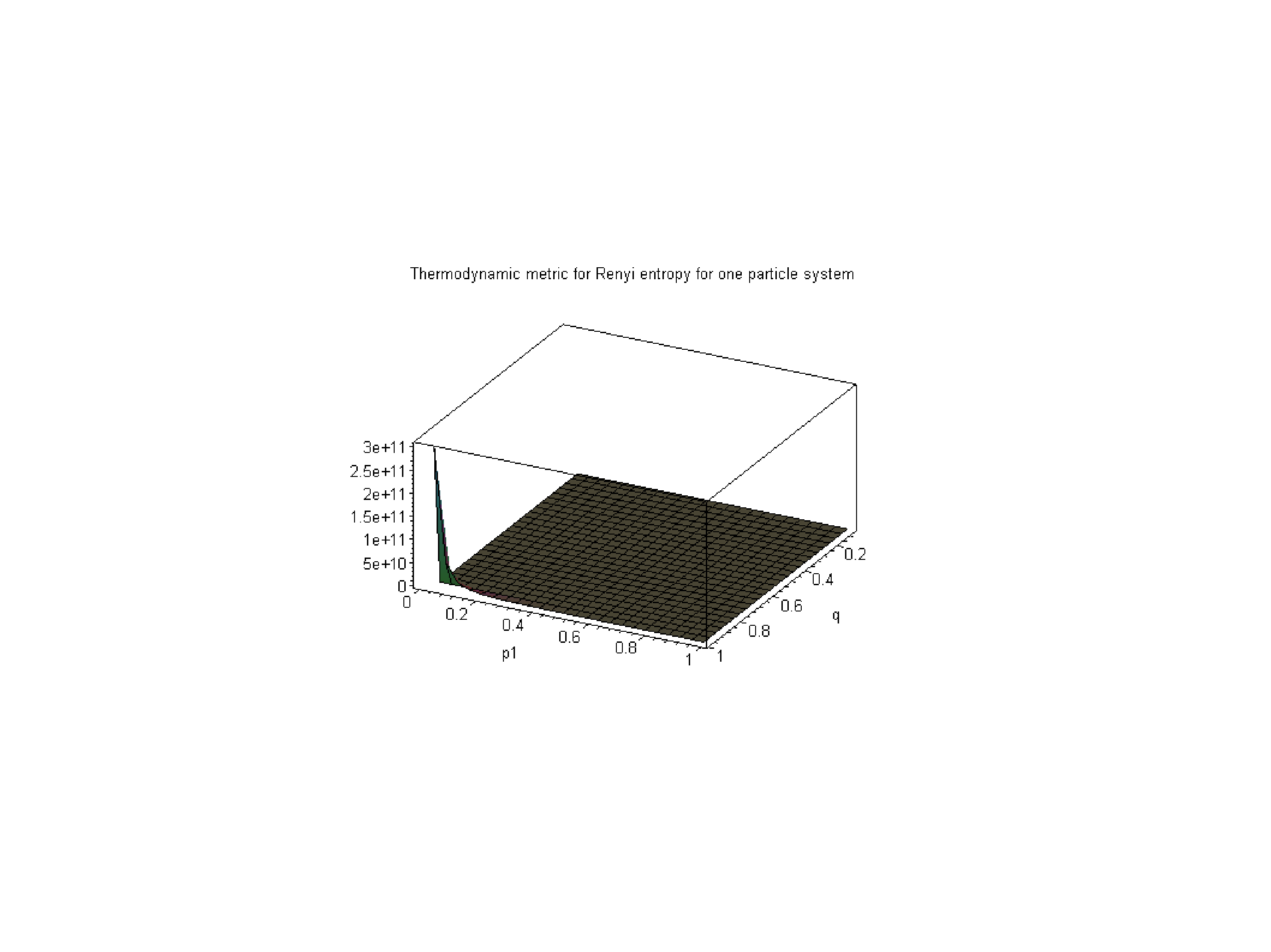}
\caption{The determinant of the metric tensor as a function of
 probability $p$,  and the entropic parameter $q$ (\ref{renyi}),  in a single
 particle R\'{e}nyi system.} \label{fig1}
\vspace*{0.5cm}
\end{figure}

\begin{figure}
\includegraphics[width=7.0cm,angle=-90]{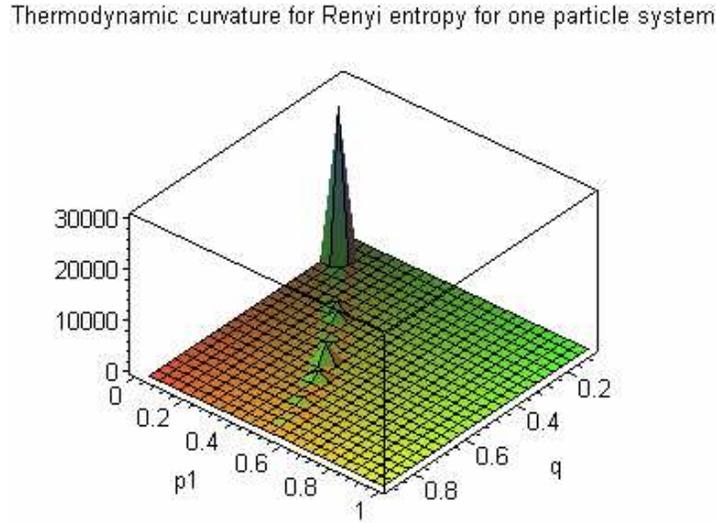}
\caption{The scalar curvature as a function of
 probability $p$,   and the entropic parameter $q$ (\ref{renyi}),  in a single
 particle R\'{e}nyi system.} \label{fig2}
\vspace*{0.5cm}
\end{figure}

Note that the extreme limit of the  system is described with the R\'{e}nyi parameter
$q: =1$; and then one finds a non-interacting configuration with $R = 0$, while
the determinant of the metric tensor $\Vert g \Vert$ diverges. The free particle
is described by the R\'{e}nyi parameter $q: = \frac{1}{3}$ (this value of $q$
corresponds to a one dimensional ideal gas  \cite{bash}).
Here,  it is easy to show that the geometric quantities reduce to their
corresponding limiting values. In particular, the scalar curvature is seen to be
\begin{equation}
R= -\frac{2}{3}\bigg(\frac{\frac{4}{9}\ln(p_1)+ \frac{2}{3}
\ln(p_1^{1/3})+ \frac{1}{3}}{(\frac{4}{9}\ln(p_1)+ \frac{2}{3}
\ln(p_1^{1/3})+ 1)^2 } \bigg),
\end{equation}
and the determinant of the metric tensor is
\begin{equation}
\Vert g \Vert = \frac{81}{16 p_1^2} \bigg(-\frac{4}{9}\ln(p_1)-
\frac{2}{3} \ln(p_1^{1/3})- 1 \bigg).
\end{equation}

The determinant of the metric (DM) and the scalar curvature for
the single particle R\'{e}nyi configuration are shown in Figs.
(\ref{fig1}) and (\ref{fig2}), respectively. From Fig.
(\ref{fig1}), it is clear that DM is non-zero and non-negative for
the chosen range of parameters and diverges  only near $p=0, q=0$.
The curvature scalar does not show any divergence in general
except at the zeros of the determinant (see Fig. (\ref{fig2})).
From fig. (\ref{fig2}),  the nature of the thermodynamic curvature
of the R\'{e}nyi entropy, that is the R\'{e}nyi correlation
length,  can be read-off for the case of free particle $q:=1/3$,
as well as for more general cases. The bumps in curvature show the
presence of non-trivial interactions in the statistical
configuration. Larger the height of a bump, stronger will be the
interaction.


\subsubsection{ Two particle R\'{e}nyi system}

From Eq. (\ref{RenyiEntropy}), we see that the expression for the entropy
of the two particle R\'{e}nyi configuration reduces to
\begin{equation}
  S_q^R( q, p_1, p_2 ):= \frac{1}{1-q} \ln( p_1^q + p_2^q ).
\end{equation}
The components of corresponding metric tensor are
\begin{eqnarray}
g_{qq}&=& \bigg(-2p_1^{2q} \ln(p_1)q+ 4 \ln(p_1^q+ p_2^q) p_1^q
p_2^q+ p_2^q \ln(p_2)^2 p_1^q  \nonumber\\&& + p_1^q \ln(p_1)^2
p_2^q + 2p_2^q \ln(p_2) p_1^q+ 2p_1^q \ln(p_1) p_2^q- 2p_2^{2q}
\ln(p_2)q \nonumber\\ && -2p_1^q \ln(p_1) p_2^q \ln(p_2)+ p_2^q
\ln(p_2)^2 q^2 p_1^q- 2p_2^{q} \ln(p_2)^2 q p_1^q\nonumber \\ && +
p_1^q \ln(p_1)^2 q^2 p_2^q- 2p_1^{q} \ln(p_1)^2 q p_2^q- 2p_2^q
\ln(p_2) q p_1^q\nonumber\\ && - 2p_1^q \ln(p_1) q p_2^q+
2p_2^{2q} \ln(p_2)+ 2p_1^{2q} \ln(p_1)+ 2 \ln(p_1^q+
p_2^q)p_1^{2q}\nonumber\\ && + 2 \ln(p_1^q+ p_2^q)p_2^{2q}- 2p_1^q
\ln(p_1) p_2^q \ln(p_2)q^2\nonumber\\ && + 4 p_1^q \ln(p_1) p_2^q
\ln(p_2)q\bigg) \times ( (p_1^q+ p_2^q)^2 (-1+q)^3 )^{-1}.
\end{eqnarray}
Explicitly, we notice that the components of the metric tensor satisfy
\begin{equation}
g_{qq}( q, p_1, p_2 )= g_{qq}( q, p_2, p_1 ).
\end{equation}
Furthermore, there exists a symmetric expression for the components of thermodynamic
metric with one $p_i$, and  in particular it is seen that
\begin{eqnarray}
g_{qp_i}&=& -\bigg( p_i^q \ln(p_i) q p_j^q- \ln(p_i) q^2 p_j^q p_i^q+ p_i^{2q}+ p_i^q p_j^q- p_j^q \ln(p_j) q p_i^q\nonumber\\
&&+ p_i^qp_j^q \ln(p_j) q^2\bigg)  \times ((-1+q)^2 p_i (p_i^q+
p_j^q)^2)^{-1},
\end{eqnarray} for $i,j= 1,2$.
Similarly,  the diagonal components of the metric tensor are
\begin{equation}
 g_{p_ip_i}= q \bigg( p_i^{q-2} q p_j^q- p_i^{2q-2}-p_i^{q-2} p_j^q \bigg) \times ((-1+q) (p_i^q+ p_j^q)^2)^{-1},
\end{equation} for $i,j= 1,2$.
Finally, the off-diagonal components turn out to be
\begin{equation}
 g_{p_ip_j}= -( p_i^{q-1} q^2 p_j^{q-1})( (-1+q) (p_i^q+ p_j^q)^2 )^{-1}, \label{2renyidistinct}
\end{equation} for  $i\neq j$.
A straightforward computation shows that the determinant of the metric tensor is
\begin{eqnarray}
\Vert g \Vert&=& -q \bigg( 2 p_1^q p_2^{3q} \ln( p_2 ) q - 2 p_1^q p_2^{3q} \ln( p_2 ) q^2
+ p_1^q p_2^{3q}+ 2 p_1^q p_2^{3q} \ln(p_1^q+ p_2^q)q+ 4 p_2^{2q}  \ln(p_1^q+ p_2^q)
 p_1^{2q} q\nonumber\\ &&+ p_2^{2q} \ln(p_2)^2q p_1^{2q}- 2 \ln(p_1)q^2 p_2^{2q} p_1^{2q}
+ 2 p_1^{2q} \ln(p_1) q p_2^{2q}- 2p_1^{2q} \ln(p_1)p_2^{2q} \ln(p_2)q+ p_1^{2q} \ln(p_1)^2
 q p_2^{2q}\nonumber\\&&+ 2p_1^{2q} \ln(p_1)p_2^{2q} \ln(p_2)q^2- 2p_1^{2q} p_2^{2q} \ln(p_2)q^2
- p_1^{2q} \ln(p_1)^2 q^2 p_2^{2q}+ 2p_2^{2q} \ln(p_2) q p_1^{2q}- p_2^{2q} \ln(p_2)^2 q^2 p_1^{2q}\nonumber\\
&&+ 2p_1^{2q} p_2^{2q}+ 2p_2^{q}  p_1^{3q} \ln(p_1) q- 2p_2^{q}
\ln(p_1) q^2 p_1^{3q}+ p_2^{q} p_1^{3q} + 2 p_2^q \ln(p_1^q+
p_2^q)p_1^{3q}q\bigg)  \times \nonumber \\ && ( p_1^2 p_2^2(p_1^q+
p_2^q)^4 (-1+q)^4 )^{-1}.
\end{eqnarray}

\begin{figure}
\includegraphics[width=7.0cm,angle=-90]{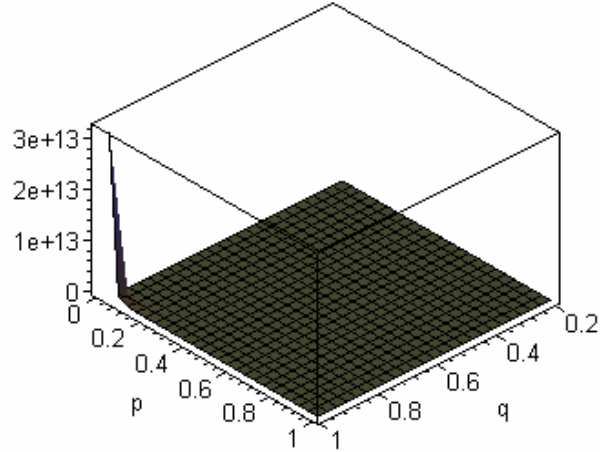}
\caption{The  determinant  of  the  metric  tensor as  a  function  of
  probability $p$ (both the particle posses equal probability),  and  the entropic parameter $q$, in a two particle R\'{e}nyi system}\label{fig3}
\vspace*{0.5cm}
\end{figure}

\begin{figure}
\includegraphics[width=7.0cm,angle=-90]{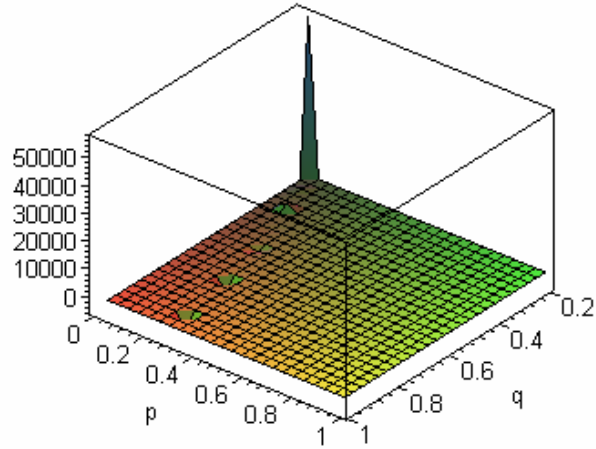}
\caption{Thermodynamic curvature as a function of probability $p$,
and the entropic parameter $q$, in a two particle R\'{e}nyi system}\label{fig4}
\vspace*{0.5cm}
\end{figure}

The scalar curvature does not have a very illuminating form, in
general. Nevertheless, for the case of equal probabilities, that
is with $p_1= p_2 = p$, the curvature is illustrated in   Fig.
(\ref{fig4}). In this case, the statistical system
corresponds to an ensemble of free particles, described by the
R\'{e}nyi parameter $ q:= 1/3 $
and the scalar curvature is
\begin{equation}
R= \frac{3}{2} \bigg(\frac{\frac{640}{81}\ln(2p^{1/3})-
\frac{1280}{243}\ln(p)- \frac{64}{9}}{(\frac{8}{3}\ln(2p^{1/3})+
\frac{16}{9}\ln(p)+ 4))^2}\bigg).
\end{equation}
From Fig. (\ref{fig3}), it is clear that DM is non-zero and non-negative for the chosen
range of parameters and  diverges  only near $p=0, q=0$.
The plots of R\'{e}nyi correlation length are shown in Fig. (\ref{fig4}) and can be easily analyzed for
(a) free  particle case with $q: =1/3$, showing no bumps, and  (b) in general,  showing bumps in some
places,  corresponding to thermodynamic interactions in the underlying system.

\subsubsection{ Three  particle R\'{e}nyi system}
In order to further understand the nature of generic R\'{e}nyi configurations,
we shall now consider a three particle system. The associated entropy, following
Eq. (\ref{RenyiEntropy}),  is
\begin{equation}
S_q^R( q, p_1, p_2 ):= \frac{1}{1-q} \ln( p_1^q + p_2^q + p_3^q).
\end{equation}
All observations made in the previous sub-section may be generalized
for arbitrary finitely many particle R\'{e}nyi configuration, and in particular,
the components of the metric tensor for the three particle system can be seen to be
\begin{eqnarray}
g_{qq}&=& \bigg(-2 p_3^{q} \ln(p_3)^2 q p_1^q+ p_3^{q} \ln(p_3)^2 q^2 p_1^q- 2 \ln(p_1) q p_1^q p_2^q\nonumber\\
&&+ 4p_1^q \ln(p_1) p_2^q \ln(p_2) q- 2p_2^q \ln(p_2) p_3^q \ln(p_3) q^2+ 2p_3^q \ln(p_3) p_2^q\nonumber\\
&&+ 2p_2^q \ln(p_2) p_3^q+ 2p_1^q \ln(p_1) p_2^q+ 2p_1^q\ln(p_1) p_3^q+ 2p_2^q \ln(p_2) p_1^q\nonumber\\
&&- 2p_1^q \ln(p_1) p_3^q \ln(p_3) q^2+ 2p_3^q \ln(p_3) p_1^q- 2p_1^q\ln(p_1) p_2^q \ln(p_2) q^2\nonumber\\
&&- 2p_3^q \ln(p_3)q p_1^q- 2p_2^q \ln(p_2) qp_1^q- 2p_1^q\ln(p_1) qp_1^q p_3^q\nonumber\\
&&- 2p_2^q \ln(p_2)q p_1^q+ p_2^q \ln(p_2)^2 q^2 p_1^q- 2\ln(p_1)^2 qp_1^q p_2^q\nonumber\\
&&+ \ln(p_1)^2 q^2 p_1^q p_3^q- 2 \ln(p_1)^2 q p_1^q p_3^q+ \ln(p_1)^2 q^2 p_1^q p_2^q+ 2p_1^{2q} \ln(p_1)\nonumber\\
&&- 2 \ln(p_1)q p_1^{2q}+ p_3^q \ln(p_3)^2 p_2^q+ p_1^{q} \ln(p_1)^2 p_2^q+ p_2^{q} \ln(p_2)^2 p_1^q\nonumber\\
&&+ p_1^q \ln(p_1)^2 p_3^q+ p_3^{q} \ln(p_3)^2 p_1^q+ p_2^q \ln(p_2)^2 p_3^q\nonumber\\
&&+ 4 \ln(p_1^q+p_2^q+p_3^q) p_1^q p_2^q- 2p_1^{q} \ln(p_1)p_2^{q} \ln(p_2)\nonumber\\
&&- 2 p_1^q \ln(p_1) p_3^q \ln(p_3)- 2p_2^q \ln(p_2) p_3^{q} \ln(p_3)\nonumber\\
&&+ 4 \ln( p_1^q + p_2^q + p_3^q) p_1^q p_3^q- 2 p_2^q \ln(p_2) q p_3^q- 2 p_3^q \ln(p_3) q p_2^q\nonumber\\
&&+ p_3^q \ln(p_3)^2 q^2 p_2^q- 2 p_3^q \ln(p_3)^2 q p_2^q+ p_2^q \ln(p_2)^2 q^2 p_3^q\nonumber\\
&&- 2 p_2^q \ln(p_2)^2 q p_3^q+ 4 p_1^q \ln(p_1) p_3^q \ln(p_3) q+ 4 p_2^q \ln(p_2) p_3^q \ln(p_3) q\nonumber\\
&&+ 2 p_3^{2q} \ln(p_3)+ 2 p_2^{2q} \ln(p_2)- 2 p_3^{2q} \ln(p_3)q- 2 p_2^{2q} \ln(p_2)q\nonumber\\
&&+ 4 \ln( p_1^q + p_2^q + p_3^q) p_2^q p_3^q+ 2 \ln( p_1^q + p_2^q + p_3^q) p_2^{2q}\nonumber\\
&&+ 2 \ln( p_1^q + p_2^q + p_3^q) p_1^{2q}+ 2 \ln(p_1^q+ p_2^q+p_3^q)p_3^{2q}\bigg) \times \nonumber\\
&&( (-1+q)^3 (p_1^q+ p_2^q+p_3^q)^2 )^{-1}.
\end{eqnarray}
We  observe that for all $i$ and $j$, the metric tensors
are pair-wise symmetric,  that is under the exchange of $\{ i,j\}$ we have
\begin{equation}
g_{qq}( q, p_i, p_j )= g_{qq}( q, p_j, p_i ).
\end{equation}
Also
\begin{eqnarray}
 g_{qp_i}&=& \bigg( -\ln(p_i) q p_i^q p_j^q- \ln(p_i) q p_i^q p_k^q
+ \ln(p_i) q^2 p_i^qp_j^q + \ln(p_i) q^2 p_i^qp_k^q\nonumber\\
&&- p_i^{2q}- p_i^q p_j^q- p_i^q p_k^q
+ p_j^q \ln(p_j) q p_i^q- p_j^q \ln(p_j) q^2 p_i^q\nonumber\\
&&+ p_k^q \ln(p_k) q p_i^q- p_k^q \ln(p_k) q^2 p_i^q\bigg) \times
((-1+q)^2 p_i (p_i^q+ p_j^q+ p_k^q)^2 )^{-1},
\end{eqnarray}
for $i,j= 1,2,3$ and $j \neq k$. The function $ g_{qp_i} $ is symmetric
under the exchange between $p_j$ and $p_k$.  Furthermore,
\begin{eqnarray}
g_{p_i p_i}&=& q \bigg( p_i^{q-2} q p_j^q+ p_i^{q-2} q p_k^q -
p_i^{2q-2}- p_i^{q-2} p_j^q- p_i^{q-2} p_k^q \bigg) \times (
(-1+q) (p_i^q+ p_j^q+ p_k^q)^2 )^{-1},
\end{eqnarray}
for $i,j= 1,2,3$ and $j \neq k $.
The functions $ g_{p_ip_i} $ are also symmetric
in $p_j$ and $p_k$.  As before,  it can be seen that for $ i\neq j$, the
distinct components of the metric tensor are
\begin{equation}
g_{p_ip_j}=
-(p_i^{q-1}q^2p_j^{q-1})((-1+q)(p_i^q+p_j^q+p_k^q)^2)^{-1}.
\label{3renyidistinct}
\end{equation}
A comparison between Eqs. (\ref{3renyidistinct}) and (\ref{2renyidistinct}) shows
that the form of the metric tensor, for the above distinct  three particle case,
is identical with that of a two particle R\'{e}nyi configuration.
In this case,  a computation for the equiprobable configurations, with
$p_1= p_2 = p_3 = p$, shows that the determinant of the metric tensor
reduces to
\begin{equation}
\Vert g \Vert= -\frac{1}{27p^6} \bigg(\frac{q^2(2 \ln(p) q - 2 q^2
\ln(p)+ 1 + 2 \ln(3p^q)q)}{(1-q)^4} \bigg).
\end{equation}
\begin{figure}
\includegraphics[width=7.0cm,angle=-90]{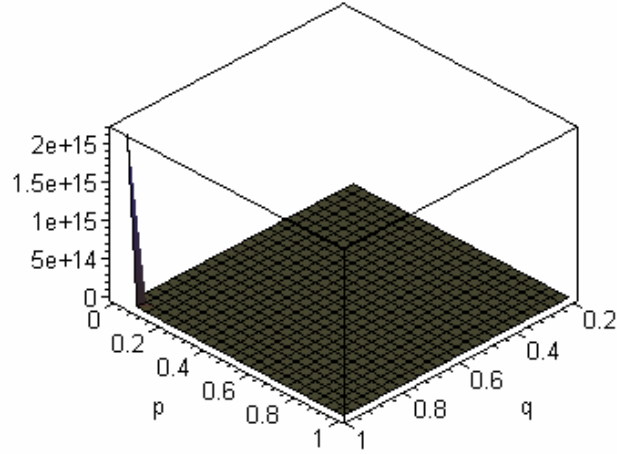}
\caption{The  determinant  of  the  metric  tensor as  a  function  of
  probability $p$,  and the entropic parameter $q$, in a three particle R\'{e}nyi system} \label{fig5}
\vspace*{0.5cm}
\end{figure}
\begin{figure}
\includegraphics[width=7.0cm,angle=-90]{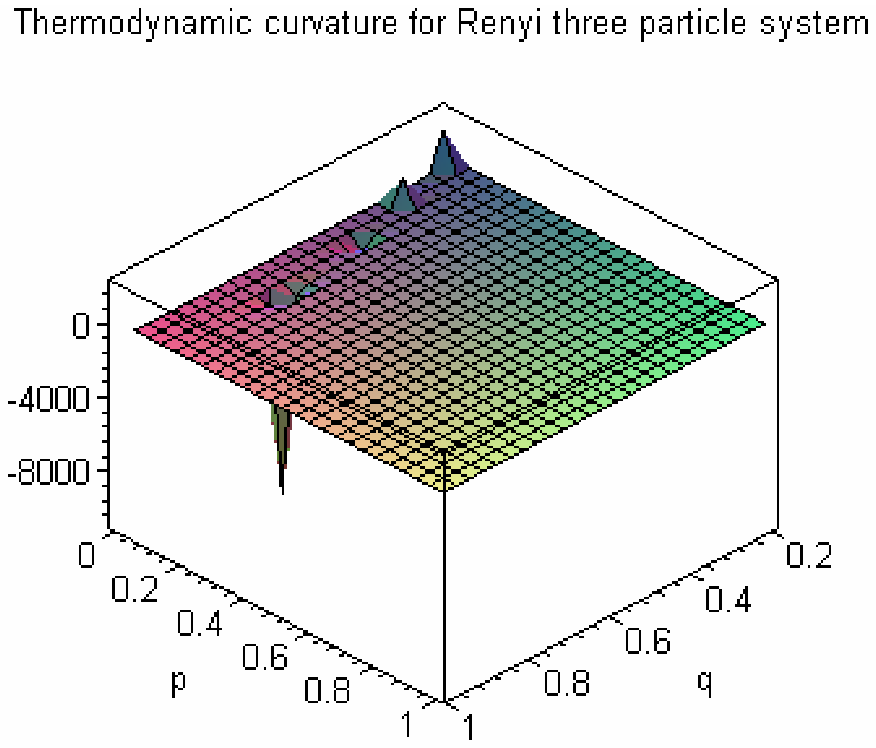}
\caption{Thermodynamic curvature as a function of probability $p$,
and the entropic parameter $q$, in a three particle R\'{e}nyi system} \label{fig6}
\vspace*{0.5cm}
\end{figure}
For non interacting R\'{e}nyi parameter $q:=1/3$; we may easily see that
the scalar curvature take the value of
\begin{eqnarray}
R&=& \frac{3}{2}\bigg(45p^{4/3}+ 20p^{4/3} \ln(3p^{1/3})+
\frac{40}{3}\ln(p)p^{4/3} + \frac{16}{9}\ln(p)^2p^{4/3}+
\frac{16}{3}\ln(p)p^{4/3}\ln(3p^{1/3})\nonumber\\
&&+ 4p^{4/3}\ln(3p^{1/3})^2\bigg)\times \bigg(4\ln(p)p^{2/3}+ 9
p^{2/3}+ 6 p^{2/3}\ln(3p^{1/3})\bigg)^{-2}.
\end{eqnarray}
Finally, the extreme value $q:=1$, leads to a degenerate metric whose
determinant has a  division by zero.  However, interestingly, we  find
in this case  a non-negative scalar curvature,
\begin{equation}
R= \frac{1}{2} \bigg( \frac{81p^4+ 324p^4\ln(3p)+
324p^4\ln(3p)^2}{(9p^2+ 18p^2\ln(3p))^2} \bigg).
\end{equation}
Here, we find  the surprising fact that the three particle
R\'{e}nyi configuration has a different geometric nature, and the
degenerate metric has an interaction. In contrast to the one and
two particle systems, the three particle system shows larger
attraction (negative curvature) (Fig. (\ref{fig6})). An
explanation of the origin of these interactions may lie in some
form of  superstatics \cite{cb01,bc03}, where the entropic
parameter $q$ is defined by physical properties of a complex
system,  that can exchange energy and heat with a thermostat. From
Fig. (\ref{fig5}), as in the previous cases, the DM is non-zero
and non-negative for the chosen range of parameters and diverges
near $(p,q)=(0,0)$.

\subsection{ Tsallis Entropy}

We now turn our attention to the Tsallis entropy \cite{ct88}. Its general form is given by \cite{cb09}
\begin{equation}
S_q^T = -\frac{1}{1-q}\Big(1 - \sum\limits_i p_i^q \Big). \label{TsallisEntropy}
\end{equation}

As in the previous subsection, we shall focus our attention on increasing number of particles.
We begin with a single particle at temperature $T$.

\subsubsection{Single particle Tsallis system}

For a single particle Tsallis configuration at the given temperature,
the entropy as a function of relative probability $p_1$ and entropic
parameter $q$ is given by
\begin{equation}
 S_q^T(q,p_1)= -\frac{1}{1-q}(1-p_1^q). \label{tsal}
\end{equation}

In this case,  the components of the covariant metric tensor are
\begin{eqnarray}
g_{qq}&=&  (-1+q)^{-3}\bigg(-2+ 2p_1^q+ 2p_1^q \ln (p_1)- 2 p_1^q
\ln (p_1)q\nonumber\\&& + p_1^q \ln (p_1)^2- 2p_1^q \ln(p_1)^2 q+ p_1^q\ln(p_1)^2 q^2\bigg), \nonumber\\
 g_{qp_1}&=&  (-1+q)^{-2} p_1^{-1+q} \bigg(-q \ln(p_1)+ \ln(p_1)q^2-1 \bigg), \nonumber\\
 g_{p_1p_1}&=&  p_1^{q-2}q.
\end{eqnarray}
The determinant of the metric tensor is
\begin{eqnarray}
\Vert g \Vert&=& -(-1+q)^{-4} p_1^2 \bigg(-2p_1^q q+ 2p_1^qq^2+ 2p_1^{2q}q
- 2p_1^{2q}q^2+ 4p_1^{2q}\ln p_1 q\nonumber\\&&- 6 p_1^{2q} \ln p_1 q^2
+ 2 p_1^{2q} \ln p_1 q^3+ p_1^{2q} \ln (p_1)^2 q- 2p_1^{2q} \ln (p_1)^2 q^2
\nonumber\\ && + p_1^{2q} \ln (p_1)^2 q^3+ p_1^{2q}\bigg).
\end{eqnarray}
The corresponding Ricci scalar is given by
\begin{eqnarray}
R &=& -\frac{1}{2} ( -1 + q )^2 \bigg( -2 q + 2 q^2 + 2 p_1^q q- 2
p_1^q q^2 + 4 p_1^q \ln( p_1 ) q- 6 p_1 q \ln( p_1 ) q^2
\nonumber\\ && + 2 p_1^q \ln( p_1 ) q^3+ p_1 q \ln( p_1 )^2 q- 2
p_1^q \ln( p_1 )^2 q^2 + p_1^q \ln( p_1 )^2 q^3 + p_1^q
\bigg)^{-2} \times \nonumber\\ && \bigg( -2- 16 q^2 + 8 q^3- 22
p_1^q \ln( p_1 ) q + 34 p_1^q \ln( p_1 ) q^2 - 20 p_1^q \ln( p_1 )
q^3\nonumber\\&&- 9 p_1^q \ln( p_1 )^2 q + 18 p_1^q \ln( p_1 )^2
q^2- 13 p_1^q \ln( p_1 )^2 q^3- 14 p_1^q q + 16 p_1^q
q^2\nonumber\\&&+ p_1^q \ln( p_1 )^2 + 4 q \ln(
p_1 )+ 4 p_1^q \ln( p_1 ) + 2 \ln( p_1 )^2 q^2- 4 \ln( p_1 )^2 q^3\nonumber\\
&&- p_1^q \ln( p_1 )^3 q + 3 p_1^q \ln( p_1 )^3 q^2- 3 p_1^q \ln(
p_1 )^3 q^3 + p_1 q \ln( p_1 )^3 q^4\nonumber\\&& + 3 p_1^q \ln(
p_1 )^2 q^4 - 2 \ln( p_1 ) q^3 + 4 p_1^q- 6 \ln( p_1 ) q^2- 8
p_1^q q^3 \nonumber\\&& + 2 \ln( p_1 )^2 q^4+ 4 \ln( p_1 ) q^4 + 8
q + 4 p_1^q q^4 \ln( p_1 ) \bigg).
\end{eqnarray}

\begin{figure}
\includegraphics[width=7.0cm,angle=-90]{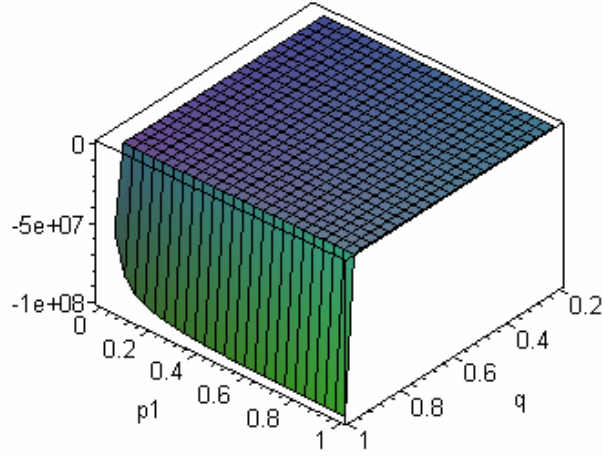}
\caption{The determinant of the metric tensor (DM) as a function of
probability $p$,  and the entropic parameter $q$, in a one particle Tsallis system.} \label{fig7}
\vspace*{0.5cm}
\end{figure}

For the extreme  value of the entropic parameter $q=1$, the underlying
statistical configuration reduces  to a non-interacting configuration, as expected,
and the metric tensor diverges as depicted in Fig. (\ref{fig7}).

\begin{figure}
\includegraphics[width=7.0cm,angle=-90]{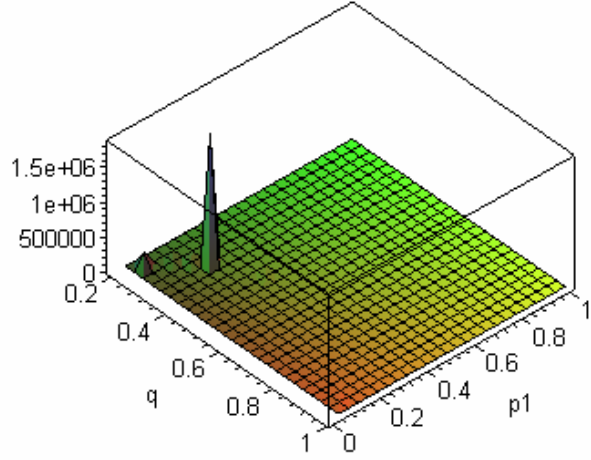}
\caption{Thermodynamic curvature as a function of probability $p_1$,
and the entropic parameter $q$,  in a one particle Tsallis system.} \label{fig8}
\vspace*{0.5cm}
\end{figure}

\subsubsection{Two particle Tsallis system}

Following Eq.(\ref{TsallisEntropy}), the Tsallis entropy of a two particle system is
\begin{equation}
S_q^T(q,p_1,p_2):=-\frac{(1-p_1^q-p_2^q)}{1-q}.
\end{equation}
The components of the thermodynamic tensor can be obtained  from
the above Tsallis entropy as
\begin{eqnarray}
 g_{qq}&=&\frac{1}{(1-q)^3} \bigg(-2+2p_1^q+2p_2^q+(2p_1^q ln(p_1)+2p_2^q ln(2p_2))(1-q)\nonumber\\
&&+(q^2-2q)( p_1^q ln(p_1)^2+  p_2^q ln(p_2)^2)+ p_1^q ln(p_1)^2+ p_2^q ln(p_2)^2\bigg), \nonumber\\
g_{q p_i}&=&\frac{ {p_i}^{q-1}}{(q-1)^2} \bigg(-q ln(p_i) + ln(p_i)q^2-1 \bigg), \nonumber\\
g_{p_i p_i}&=&  p_{i}^{q-2}q.
\end{eqnarray}
As in the case of  R\'{e}nyi entropy,
we see in this case  similar symmetries between the components of the metric tensor.
The determinant of the metric tensor is
\begin{eqnarray}
\Vert g \Vert&=&  q (1-q)^{-4} \bigg( 2 p_2^{q-2} q^3
p_1^{2q-2}\ln(p_1)+ p_2^{2q-2} q^3 \ln(p_2)^2 p_1^{q-2}
\nonumber\\ &&+ p_2^{q-2} q^3 p_1^{2q-2} \ln( p_1 )^2 +
2p_2^{2q-2} q^3 \ln( p_2 ) p_1^{q-2} -p_2^{2q-2}q^2
p_1^{q-2}\nonumber\\ &&- 6 p_2^{q-2} q^2 p_1^{2q-2} \ln( p_1 )
- 2 p_2^{q-2} q^2 p_1^{2q-2}+ 2p_2^{q-2} q^2 p_1^{q-2}\nonumber\\
&&- 2p_2^{q-2} q^2 p_1^{2q-2}\ln( p_1 )^2 - 6 p_2^{2q-2} q^2 \ln(
p_2 ) p_1^{q-2}- 2 p_2^{2q-2} q^2 p_1^{q-2} \ln( p_2 )^2\nonumber\\
&&+ 2p_2^{q-2} q p_1^{2q-2}+ p_2^{q-2} q p_1^{2q-2} \ln( p_1 )^2
+ 4 p_2^{2q-2} q  \ln( p_2 ) p_1^{q-2}\nonumber\\
&&+ 4 p_2^{q-2} q  p_1^{2q-2} \ln( p_1 ) + p_2^{2q-2} q \ln( p_2
)^2 p_1^{q-2}+ 2p_2^{2q-2} q  p_1^{q-2}\nonumber\\
&&- 2p_2^{q-2} q  p_1^{q-2}+ p_2^{2q-2} p_1^{q-2} + p_1^{2q-2}
p_2^{q-2} \bigg) .
\end{eqnarray}

\begin{figure}
\includegraphics[width=7.0cm,angle=-90]{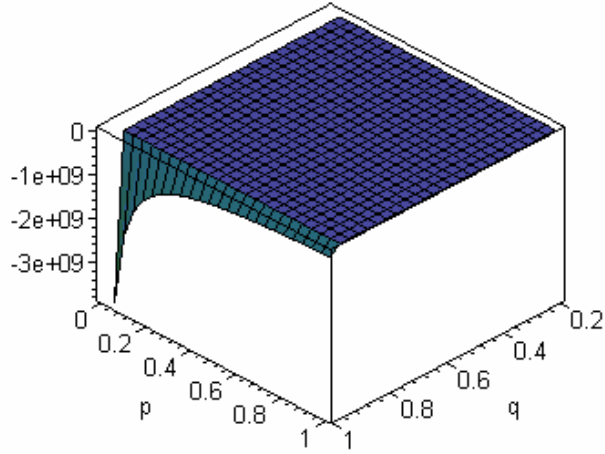}
\caption{The  determinant  of  the  metric  tensor as  a  function  of
  probability $p$,  and the entropic parameter $q$, in a two particle Tsallis system.} \label{fig9}
\vspace*{0.5cm}
\end{figure}

\begin{figure}
\includegraphics[width=7.0cm,angle=-90]{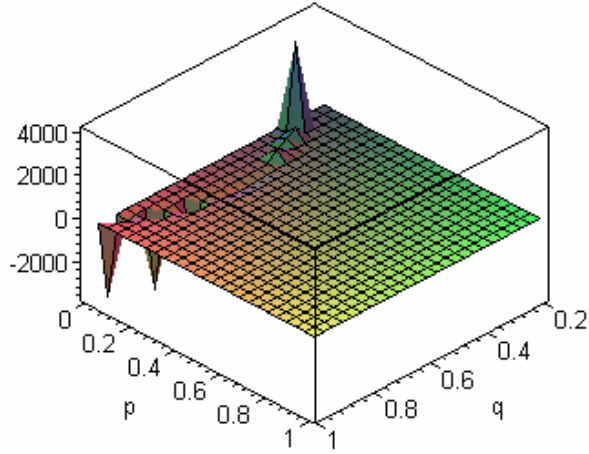}
\caption{Thermodynamic curvature as a function of probability  $p$,
and the entropic parameter $q$, in a two particle Tsallis system.} \label{fig10}
\vspace*{0.5cm}
\end{figure}
For the case of $q=1/3$, corresponding to the case of free
particles \cite{bash}, and  equal probabilities $p_1=p_2=p$, the
thermodynamic metric and curvature, in the case of two particle
Tsallis system have the following forms
\begin{eqnarray}
\Vert g \Vert= -\frac{27}{16} \bigg( \frac{40 \ln(p)}{27 p^3}
+\frac{8 \ln(p)^2}{27 p^3}+\frac{26}{9 p^3}-\frac{4}{9
p^{10/3}}\bigg)\\,
\end{eqnarray}
and
\begin{eqnarray}
R&=&-\frac{3}{2}\bigg(-\frac{8}{27} \ln(p)^2 p^{1/3}+
\frac{40}{27}p^{1/3}\ln(p)- \frac{4}{9} + \frac{26}{9}
p^{1/3}\bigg)^{-2}  \times \nonumber\\&&
\bigg(\frac{944}{2187}\ln(p)+ \frac{112}{2187}\ln(p)^2-
\frac{64}{6561}p^{1/3}\ln(p)^4-
\frac{1216}{6561}p^{1/3}\ln(p)^3\nonumber\\&&-
\frac{5392}{6561}\ln(p)^2 p^{1/3}- \frac{928}{729} p^{1/3}\ln(p)-
\frac{616}{729}p^{1/3}+ \frac{308}{729}\bigg).
\end{eqnarray}
These are plotted in Figs. (\ref{fig9}) and (\ref{fig10}),
respectively, for different values of $q$. For  $q=1$,  the
underlying configuration is seen to possess $R=0$, while the
determinant of the metric tensor $\Vert g \Vert$ is, again, seen
to diverge.  From Fig.   (\ref{fig10}) for  the  correlation
length (thermodynamic curvature), the nature of statistical
interactions show interesting features, in the form of bumps, near
the limit $q=1$.

\subsubsection{Three particle Tsallis system}

\begin{figure}
\includegraphics[width=7.0cm,angle=-90]{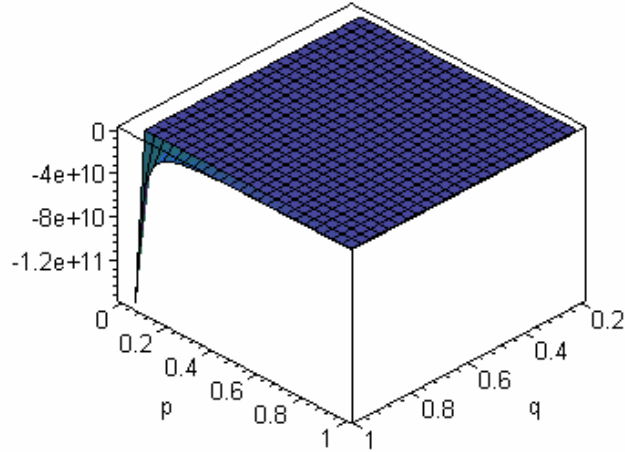}
\caption{The  determinant  of  the  metric  tensor as  a  function  of
probability $p$,  and the entropic parameter $q$, in a three particle Tsallis system.} \label{fig11}
\vspace*{0.5cm}
\end{figure}

\begin{figure}
\includegraphics[width=7.0cm,angle=-90]{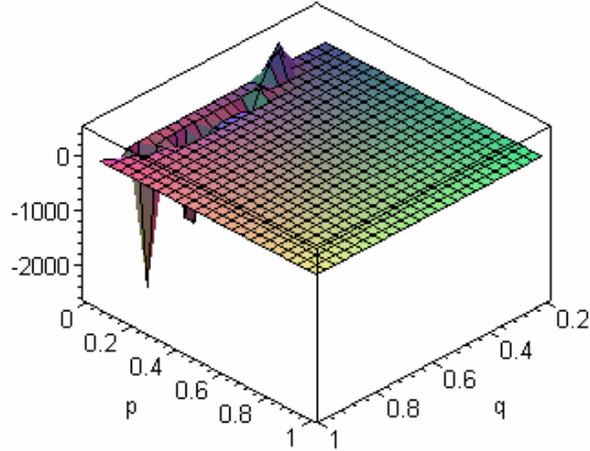}
\caption{Thermodynamic curvature as a function of the probability $p$,
and the entropic parameter $q$, in a three particle Tsallis system.} \label{fig12}
\vspace*{0.5cm}
\end{figure}

We  now consider the three particle system whose Tsallis entropy is
\begin{equation}
S_q^T( q, p_1, p_2, p_3 )= \frac{1}{1-q} ( 1- p_1^q- p_2^q- p_3^q ).
\end{equation}
The components of the covariant metric tensor are
\begin{eqnarray}
g_{qq}& =&  ( -1 + q )^{- 3} \bigg(-2 + 2 p_1^q + 2 p_2^q + 2
p_3^q + 2 p_1^q \ln(p_1) - 2 p_1^q \ln(p_1) q + 2 p_2^q
\ln(p_2)\nonumber\\ &&- 2 p_2^q \ln( p_2 ) q + 2 p_3^q \ln( p_3 )-
2 p_3^q \ln( p_3 ) q + p_1^q \ln( p_1 )^2- 2 p_1^q \ln( p_1 )^2 q
\nonumber\\&&+ p_1^q \ln( p_1)^2 q^2 + p_2^q \ln( p_2 )^2- 2 p_2^q
\ln( p_2 )^2 q + p_2^q \ln( p_2 )^2 q^2 + p_3^q \ln( p_3
)^2\nonumber\\ &&- 2 p_3^q \ln( p_3 )^2 q
+ p_3^q \ln( p_3 )^2 q^2 \bigg),\nonumber\\
g_{qp_i} &=&  ( -1 + q )^{- 2}  p_i^{q-1} \bigg(-\ln( p_i ) q + \ln( p_i ) q^2- 1 \bigg); i= 1-3, \nonumber\\
g_{p_ip_i} &=& p_i^{q-2}q.
\end{eqnarray}
In this case,  symmetries similar to the ones  seen before, are
observed. The DM for the equiprobable configuration, with $p_i =
p$, reduces to
\begin{eqnarray}
\Vert g \Vert&=& -q^2p^{3 q-6} (-1 + q )^{- 4}\bigg( 2 q^2- 18 p^q
\ln( p ) q^2 + 3 p^q \ln( p )^2 q^3 + 6 p^q \ln( p ) q^3
\nonumber\\ && + 12 p^q \ln( p ) q + 3 p^q \ln( p )^2 q - 6 p^q
\ln( p )^2 q^2 + 3 p^q + 6 p^q q - 6 p^q q^2- 2 q \bigg).
\end{eqnarray}
This is shown in Fig. (\ref{fig11}).  In this limit,
we see that the expressions for the scalar curvature,  for $q=1/3$, simplifies to
\begin{eqnarray}
R&=&-\frac{3}{2}\bigg( -\frac{4}{9}+ \frac{20}{9}p^{1/3}\ln(p)+
\frac{4}{9}p^{1/3}\ln (p)^2+ \frac{13}{3} p^{1/3} \bigg)^{-2}
\times \bigg(\frac{736}{729}\ln(p)\nonumber\\&&-
\frac{760}{81}p^{1/3}\ln(p)- \frac{9520}{2187}p^{1/3}\ln(p)^2+
\frac{44}{27}- \frac{112}{2187}\ln(p)^4 p^{1/3} \nonumber\\&& -
\frac{1888}{2187} \ln(p)^3p^{1/3}-\frac{253}{27}p^{1/3}+
\frac{80}{729}\ln(p)^2\bigg).
\end{eqnarray}
Finally, at  $q= 1$, we find that the scalar curvature reduces to
$R= \frac{1}{6p}$, while $\Vert g \Vert$ diverges. Consistent with
the three particle R\'{e}nyi case, here again, we find that the
three particle Tsallis system has a different nature than that of
the two particle one. The DM and thermodynamic curvature are shown
in  Figs.  (\ref{fig11}) and (\ref{fig12}), respectively.  In
particular, it turns out again that there exists a degenerate
metric tensor, and the underlying configuration has thermodynamic
interactions, as can be inferred from the bumps in the plot of
thermodynamic curvature,  Fig.  (\ref{fig12}). An explanation of
the origin of these interactions may, again, lie in some form of
superstatics \cite{cb01,bc03}, where the entropic parameter $q$ is
defined by physical properties of the  system,  and is thus a
signature of a complex system.

The DM and the scalar curvature for the one particle Tsallis
configuration are shown in Figs. (\ref{fig7}) and (\ref{fig8}),
respectively. From Fig. (\ref{fig7}), it is clear that DM is
non-zero and acquires  large-negative  values near $q=1$. As
expected, the curvature scalar does not show any divergence in
general except at the zeros of the determinant of the metric
tensor, as seen in Fig. (\ref{fig8}), and the peaks always acquire
a positive value in the chosen domain of parameters. The bumps in
curvature show the presence of non-trivial interactions in the
statistical configuration. Larger the height of a bump, stronger
will be the interactions. In the case of a two particle
configuration,  the system is seen to become more stable, except
for $p=0$ and $q=1$, as seen in Fig.  (\ref{fig9}). The curvature
scalar, Fig.  (\ref{fig10}),  shows both positive and negative
values, which implies highly non-trivial interaction present in
the system depending upon the parameter space $(p, q)$. Similar
conclusions hold for the three particle Tsallis configurations,
Figs.  (\ref{fig11}) and (\ref{fig12}).

\subsection{Abe Entropy}
Now, we consider the Abe entropy \cite{cb09,sa97}, which
is a symmetric  modification of the Tsallis entropy, an inspiration from
the theory of quantum groups.
It is given by
\begin{equation} \label{AbeEntropy}
S_q^{Abe} = - \sum\limits_{i}\frac{p_i^{q} - p_i^{q^{-1}}}{q - q^{-1}}. \label{abe}
\end{equation}
It is related to the Tsallis entropy by
\begin{equation}
S_q^{Abe} = \frac{(q-1)S_q^T - (q^{-1}-1)S_{q-1}^T}{q - q^{-1}}.
\end{equation}

In the subsequent analysis, we focus our attention on systems,
described by the Abe entropy.

\subsubsection{Single particle Abe system}
From the Eq. (\ref{AbeEntropy}),  the entropy of a single
particle Abe system is given by
\begin{equation}
S(q,p_1)= -\frac{p_1^q- p_1^{1/q}}{q-1/q}.
\end{equation}
The components of the covariant metric tensor are
\begin{eqnarray}
g_{qq}&=& q^{-3}(q^2-1)^{-3}\bigg (p_1^q \ln (p_1)^2q^8- 2p_1^q
\ln (p_1)^2 q^6\nonumber\\&&+ p_1^q \ln (p_1)^2q^4 - p_1^{1/q} \ln
(p_1)^2 q^4 + 2p_1^{1/q} \ln (p_1)^2 q^2\nonumber\\&& - p_1^{1/q}
\ln (p_1)^2- 4p_1^{1/q} \ln (p_1) q^5+ 4p_1^{1/q} \ln (p_1)
q^3\nonumber\\&& - 2p_1^q \ln (p_1) q^7+ 2p_1^q \ln (p_1) q^3+
2p_1^q q^6+ 6p_1^qq^4 \nonumber\\&&- 2q^6p_1^{1/q}- 6p_1^{1/q}
q^4\bigg), \nonumber\\ g_{qp_1}&=& q^{-2}(q^2-1)^{-2}\bigg
(p_1^{q-1} \ln (p_1) q^6- p_1^{q-1} \ln (p_1) q^4- 2p_1^{q-1}
q^3\nonumber\\&& + p_1^{-1+1/q} \ln (p_1) q^2 -p_1^{-1+1/q}
\ln (p_1)+ 2p_1^{-1+1/q} q^3\bigg),  \nonumber \\
g_{p_1p_1}&=& q^{-1}(q+1)^{-1} \bigg(p_1^{q-2}q^3
+p_1^{-2+1/q}\bigg).
\end{eqnarray}
The determinant of the metric tensor is
\begin{eqnarray}
\Vert g \Vert &=& - q^{-3}p_1^{-2}(q^2-1)^{-4}\bigg(4
\ln(p_1)p_1^{2/q}q^5- 4\ln(p_1)p_1^{2/q}q^3 \nonumber\\&&-
8p_1^{q+1/q} \ln(p_1)^2 q^5 +4p_1^{q+1/q}\ln(p_1)^2q^7-
p_1^{q+1/q} \ln(p_1)^2q^2\nonumber\\&&+ 6p_1^{2/q}q^4 +
p_1^{2/q}\ln(p_1)^2-p_1^{2q} \ln(p_1)q^9+
2\ln(p_1)q^{10}p_1^{2q}\nonumber\\&&+ 4p_1^{q+1/q}\ln(p_1)^2q^3 -
6p_1^{2q}q^7+p_1^{q+1/q} \ln(p_1)^2q^4-
4p_1^{2q}\ln(p_1)q^8\nonumber\\&&+ 2q^6p_1^{2/q} +
p_1^{2q}\ln(p_1)^2q^6+p_1^{q+1/q} \ln(p_1)^2q^6+
6p_1^{q+1/q}q^3\nonumber\\&& + 6p_1^{2q}q^6+
2p_1^{2/q}q^5+4p_1^{2q}q^5-6p_1^{2/q}q^3-
2p_1^{2q}q^9\nonumber\\&& + 2p_1^{2q}q^8+ 2\ln(p_1)q^5p_1^{2q}-
2p_1^{2q}\ln(p_1)^2q^8+2 \ln(p_1) q^6 p_1^{2q}\nonumber\\&& +
p_1^{2/q} \ln(p_1)^2- 2p_1^{2/q} \ln(p_1)^2q^2+ p_1^{2q}
\ln(p_1)^2q^{10}\nonumber\\&& + 2q^9 p_1^{q+1/q}- 2p_1^{q+1/q}q^8-
8p_1^{q+1/q}q^6+ 6p_1^{q+1/q}q^7\nonumber\\&& -
6p_1^{q+1/q}q^4-6p_1^{q+1/q}q^5- p_1^{q+1/q} \ln(p_1)^2q^8+
4p_1^{q+1/q}q^5 \ln(p_1)\nonumber\\&& +6\ln(p_1)q^2p_1^{q+1/q}+
8\ln(p_1)p_1^{q+1/q}q^8- 2\ln(p_1)p_1^{q+1/q}q^7\nonumber\\&&
-10\ln(p_1)p_1^{q+1/q}q^6- 4\ln(p_1)q^4p_1^{q+1/q}-
2\ln(p_1)q^3p_1^{q+1/q}\bigg).
\end{eqnarray}

\begin{figure}
\includegraphics[width=7.0cm]{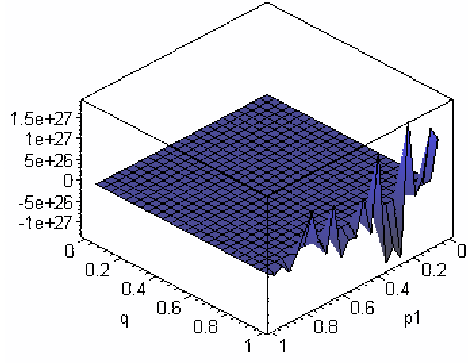}
\caption{The  determinant  of  the  metric  tensor as  a  function  of
  probability  $p_1$, and entropic parameter $q$, in  a single
  particle Abe system} \label{fig13}
\vspace*{0.5cm}
\end{figure}

\begin{figure}
\includegraphics[width=7.0cm]{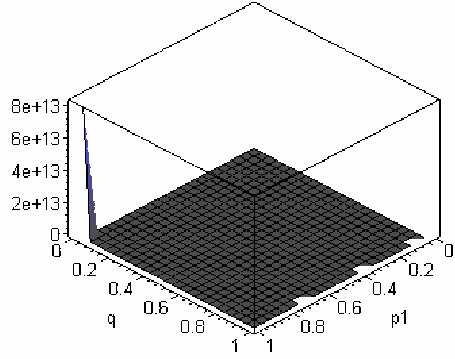}
\caption{Scalar curvature as a function of $p_1$,  and entropic parameter $q$,  in a single particle Abe system.} \label{fig14}
\vspace*{0.5cm}
\end{figure}

It can be shown  that the thermodynamic scalar curvature for $q= 1/3$ is given by
\begin{equation}
R= \frac{4}{9p_1^4 g^2(p_1)}\bigg(\frac{4096}{177147} \bigg)^2
N^{(1)}(p_1),
\end{equation}
where the numerator function $N^{(1)}(p_1)$ is
\begin{eqnarray}
N^{(1)}(p_1)&=& -\frac{39680}{129140163}p_1
-\frac{655360}{129140163}p_1^{11/3} \ln(p_1)^4\nonumber\\&&
-\frac{1863680}{43046721}p_1^{11/3} \ln(p_1)^3
-\frac{8270848}{387420489} p_1^{11/3} \ln(p_1)^2\nonumber\\&&
+\frac{18558208}{387420489}p_1^{11/3} \ln(p_1)
+\frac{61952}{177147}p_1^9\nonumber\\&&
-\frac{5771264}{4782969}p_1^{19/3} \ln(p_1)^3
-\frac{27474944}{14348907} p_1^{11/3} \ln(p_1)^2\nonumber\\&&
-\frac{19625216}{14348907} p_1^{19/3} \ln(p_1)
+\frac{118528}{387420489} p_1 \ln(p_1)\nonumber\\&&
+\frac{5120}{43046721}p_1 \ln(p_1)^2 + \frac{4096}{387420489} p_1
\ln(p_1)^3\nonumber\\&& -\frac{4096}{6561} p_1^9 \ln(p_1)^3
+\frac{10496}{19683} \ln(p_1) p_1^9\nonumber\\&&
+\frac{7165952}{129140163}p_1^{11/3} -\frac{1936640}{4782969}
p_1^{19/3}\nonumber\\&& -\frac{1024}{2187} \ln(p_1)^2 p_1^9.
\end{eqnarray}
The determinant of the metric tensor is given by
\begin{eqnarray}
\Vert g \Vert &=& -\frac{177147}{4096 p_1^2}
\bigg(\frac{64}{81}p_1^6 \ln(p_1)^2 +\frac{64}{59049}p_1^{2/3}
\ln(p_1)^2 -\frac{32}{243}p_1^6 \ln(p_1)\nonumber\\&&
+\frac{3584}{6561}p_1^{10/3} \ln(p_1)
+\frac{128}{6561}p_1^{10/3}\ln(p_1)^2 +\frac{2264}{19683}
p_1^{10/3}\nonumber\\&& +\frac{608}{59049} p_1^{2/3} \ln(p_1)
+\frac{436}{19683}p_1^{2/3}-\frac{100}{729} p_1^6 \bigg).
\end{eqnarray}

\subsubsection{Two particle Abe system}
Similarly from  Eq. (\ref{AbeEntropy}),  the entropy of the two
particle Abe system is
\begin{equation}
S(q,p_1,p_2)= -\frac{p_1^q- p_1^{1/q}}{q-1/q} -\frac{p_2^q- p_2^{1/q}}{q-1/q}.
\end{equation}
The components of the covariant metric tensor are
\begin{eqnarray}
g_{qq}&=& q^{-3}(q^2-1)^{-3}
\bigg(p_1^q \ln (p_1)^2q^4+ 2p_2^q \ln (p_2) q^3- p_2^{1/q} \ln (p_2)^2q^4\nonumber\\&&
 -2p_1^q \ln (p_1) q^7+ 2p_1^q \ln (p_1) q^3 +4 p_1^{1/q} \ln (p_1) q^3 +2p_1^{1/q} \ln (p_1)^2 q^2
\nonumber\\&&
-2p_1^q \ln (p_1)^2 q^6
-p_1^{1/q} \ln (p_1)^2 q^4+ p_2^q \ln (p_2)^2 q^4 -2 \ln (p_2) p_2^q q^7\nonumber\\&&
 -2p_2^q \ln (p_2)^2 q^6
-4 \ln (p_2)q^5p_2^{1/q} +2p_2^{1/q} \ln (p_2)^2q^2 +4p_2^{1/q} \ln (p_2)q^3\nonumber\\&&
 +p_1^q \ln (p_1)^2 q^8
-4 \ln (p_1) p_1^{1/q} q^5 + p_2^q \ln (p_2)^2 q^8 -2p_2^{1/q}q^6\nonumber\\&&
 +2p_2^q q^6 -6p_2^{1/q}q^4
-2p_1^{1/q} q^6+ 6p_1^q q^4 +2 p_1^q q^6 -6p_1^{1/q} q^4 \nonumber\\&&
-p_2^{1/q}\ln (p_2)^2 -p_1^{1/q} \ln (p_1)^2
+6p_2^q q^4\bigg), \nonumber \\
g_{qp_i}&=& q^{-2}(q^2-1)^{-2}\bigg(p_i^{q-1} \ln (p_i) q^6-
p_i^{q-1} \ln (p_i) q^4- 2p_i^{q-1} q^3\nonumber\\&& +
p_i^{-1+1/q} \ln (p_i) q^2
-p_i^{-1+1/q} \ln (p_i)+ 2p_i^{-1+1/q} q^3\bigg), \ i= \ 1,2 ,\nonumber \\
g_{p_ip_i}&=& q^{-1}(q+1)^{-1} \bigg(p_i^{q-2}q^3+p_i^{-2+1/q}\bigg).
\end{eqnarray}

\begin{figure}
\includegraphics[width=7.0cm]{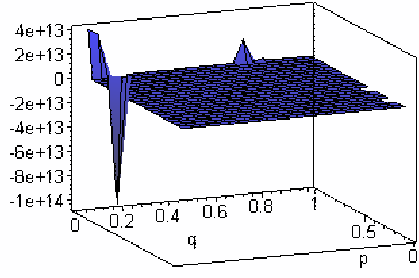}
\caption{The  determinant  of  the  metric  tensor as  a  function  of
  probability $p$, and entropic parameter $q$,   in  a two  particle Abe system.} \label{fig15}
\vspace*{0.5cm}
\end{figure}

\begin{figure}
\includegraphics[width=7.0cm]{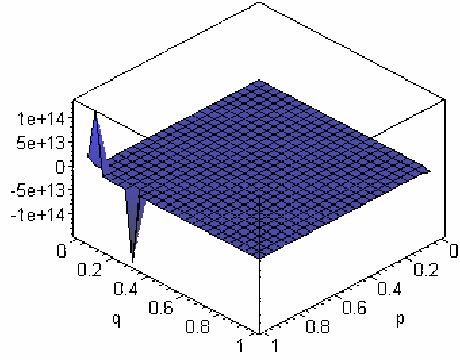}
\caption{Scalar curvature as a function of $p$ and $q$, in a two particle Abe system.} \label{fig16}
\vspace*{0.5cm}
\end{figure}

For the case of $q=1/3$, and  equal values of the probabilities $p_i=p$,
the determinant of the metric tensor is
\begin{eqnarray}
\Vert g \Vert  &=& -\frac{1594323}{16384 p^4}
\bigg(\frac{872}{531441}p +\frac{640}{6561}p^{19/3} \ln(p)^2
-\frac{1216}{1594323}p \ln(p)\nonumber\\&& +\frac{128}{81}p^9
\ln(p)^2 +\frac{128}{1594323}p \ln(p)^2 -\frac{64}{243} p^9
\ln(p)\nonumber\\&& +\frac{10816}{177147} p^{11/3} \ln(p)
+\frac{640}{177147}p^{11/3}\ln(p)^2 +\frac{2368}{2187} p^{19/3}
\ln(p)\nonumber\\&& +\frac{4328}{19683}p^{19/3}
+\frac{28072}{531441}p^{11/3}-\frac{200}{729} p^9 \bigg),
\end{eqnarray}
and the associated scalar curvature is
\begin{equation}
R= \frac{3}{2p^8 g^2(p)} \bigg(\frac{1594323}{16384} \bigg)^2
N^{(2)}(p).
\end{equation}
Here, we notice that the function $N^{(2)}(p)$ may, intriguingly, be expressed as
\begin{equation}
N^{(2)}(p)= \sum_{a\in A} \sum_{b \in B} \alpha_{ab} \ln(p)^a p^b. \label{N2}
\end{equation}
We observe further that the respective indices sets are defined as
\begin{eqnarray}
A&=&\{0,1,2,3,4 \}, \nonumber \\
B&=&\{\ 15, \frac{37}{3}, \frac{29}{3}, 7,\frac{13}{3},\frac{5}{3} \}.\nonumber \\
\end{eqnarray}

\begin{figure}
\includegraphics[width=7.0cm]{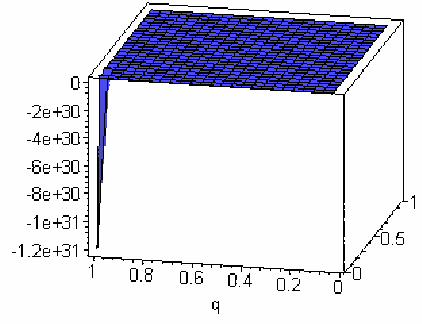}
\caption{The  determinant  of  the  metric  tensor as  a  function  of
  probability  $p$, and entropic parameter $q$,  in  a three  particle Abe system.} \label{fig17}
\vspace*{0.5cm}
\end{figure}

\begin{figure}
\includegraphics[width=7.0cm]{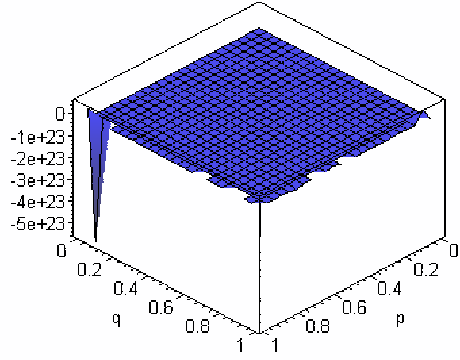}
\caption{Scalar curvature as a function of parameters $p$ and $q$, in a three particle Abe system.} \label{fig18}
\vspace*{0.5cm}
\end{figure}

\subsubsection{Three particle Abe system}
The entropy of the three
particle Abe system,   is given by (Eq. (\ref{AbeEntropy}))
\begin{equation}
S(q,p_1,p_2,p_3,p_4)=-\frac{p^q_1-p_1^{1/q}}{q-1/q}
-\frac{p^q_2-p_2^{1/q}}{q-1/q}-\frac{p^q_3-p_3^{1/q}}{q-1/q}.
\end{equation}
The component of covariant thermodynamic metric tensor in this case are
\begin{eqnarray}
g_{q,q}&=& \bigg(-2\ln(p_3)q^7 p_3^q+4\ln(p_3)q^3 p_3^{1/q}+2\ln(p_3)q^3 p_3^{q}+6p_2^q q^4-p_1^{(1/q)}\ln(p_1)^2\nonumber\\
&&-2q^6 p_3^{1/q}-p_2^{1/q}\ln(p_2)^2+p_1^q \ln(p_1)^2q ^8-2p_1^q \ln(p_1)^2q ^6-p_1^{1/q}\ln(p_1)^2 q^4\nonumber\\
&&+ p_3^q \ln(p_3)^2 q^4 +p_2^q\ln(p_2)^2 q^4 +p_1^q\ln(p_1)^2 q^4 +4\ln(p_2)p_2^{1/q} q^3\nonumber\\
&&+ 2\ln(p_2) q^3 p_2^q-2\ln(p_1) q^7 p_1^q +2 p_1^{1/q}\ln(p_1)^2 q^2-4\ln(p_1)p_1^{1/q} q^5\nonumber\\
&&+4\ln(p_1)p^{1/q} q^3 +2 \ln(p_1) q^3 p_1^q+p_2^q\ln(p_2)^2 q^8-p_2^{1/q}\ln(p_2)^2 q^4\nonumber\\
&& + 2p_2^{1/q}\ln(p_2)^2 q^2-4\ln(p_2)p_2^{1/q} q^5-2p_2^q\ln(p^2)^2 q^6-2p_3^q\ln(p_3)^2 q^6\nonumber\\
&&-p_3^{1/q}\ln(p_3)^2 q^4 +2 p_3^{1/q}\ln(p_3)^2 q^2-4\ln(p_3)p_3^{1/q} q^5 +p_3^q\ln(p_3)^2 q^8\nonumber\\
&& -2\ln(p_2) q^7 p_2^q + 2 p_3^q q^6 -6 p_3^{1/q} q^4 -2 q^6 p_2^{1/q}+2 p_2^q q^6-6p_2^{1/q} q^4\nonumber\\
&&-2q^6 p_1^{1/q} +6 p_1^q q^4 +2p_1^q q^6 -6p_1^{1/q} q^4 +6p_3^q
q^4 -p_3^{1/q}\ln(p_3)^2\bigg)\times  \nonumber\\
&& (q^{-3}(q^2-1)^{-3}),\nonumber\\
g_{q,p_i}&=&\bigg(p_i^{q-1}(\ln(p_i)q^6-\ln(p_i)q^4-2 q^3)
+p_i^{-1+1/q}(\ln(p_i)q^2 \nonumber\\ &&-\ln(p_i)+2 q^3)\bigg) \times (q^2(q^2-1)^2), \nonumber\\
g_{p_i,p_i}&=& \bigg(p_i^{q-2}q^3 +p_i^{-2+1/q}\bigg) \times
(q(q+1)).
\end{eqnarray}

The determinant of the metric tensor for $p_i\equiv p$ and $q=1/3$
has a simple form,

\begin{eqnarray}
\Vert g \Vert &=&-\frac{14348907}{65536 p^6} \bigg(p^{4/3}(a_1+a_2
\ln(p)+a_3 \ln(p)^2) + p^{4}(b_1+b_2 \ln(p)+b_3 \ln(p)^2)
\nonumber\\ &&+p^{20/3}(c_1+c_2 \ln(p)+c_3 \ln(p)^2)
+p^{28/3}(d_1+d_2 \ln(p)+d_3 \ln(p)^2)\nonumber\\
&&+p^{12}(e_1+e_2 \ln(p)+e_3 \ln(p)^2 \bigg),
\end{eqnarray}
where  the coefficients  $a_{i}$, $b_{i}$------$e_{i}$'s are
\begin{eqnarray}
a_1&=& \frac{436}{4782969}, \ \ a_2=\frac{608}{14348907},  \ \ a_3= \frac{64}{14348907}; \nonumber \\
b_1&=& \frac{25808}{4782969},  \ \ b_2= \frac{7232}{1594323},  \ \ b_3= \frac{512}{1594323}; \nonumber \\
c_1&=& \frac{200}{2187},  \ \ c_2= \frac{8960}{59049},  \ \  c_3= \frac{640}{59049}; \nonumber\\
d_1&=& \frac{688}{2187},  \ \ d_2= \frac{3520}{2187},  \ \ d_3= \frac{512}{2187}; \nonumber \\
e_1&=& -\frac{100}{243}, \ \  e_2= -\frac{32}{81},  \ \ e_3=
\frac{64}{27}.
\end{eqnarray}


We find that the determinant of the metric tensor factorizes, and
the scalar curvature may be expressed as
\begin{equation}
R(p)= -\frac{1}{2}  \frac{N^{(3)}(p)}{g_1(p)g_2(p)g_3(p)}.
\end{equation}
Here, the numerator function $N^{(3)}(p)$ may again be expressed
as earlier for the case of a two particle Abe system (\ref{N2}),
except  that the index sets $A$ and $B$ are now defined as
\begin{eqnarray}
A&=&\{0,1,2,3,4,5,6 \}; \nonumber \\
B&=&\{\ 24, \frac{64}{3}, \frac{56}{3}, 16, \frac{40}{3}, \frac{32}{3}, \frac{16}{3}, 8, \frac{8}{3} \} .
\end{eqnarray}
The factors appearing in the denominator of the scalar curvature are
\begin{eqnarray}
g_1(p) &=& -\frac{100}{729}p^6+ \frac{2264}{19683}p^{10/3}+ \frac{64}{81} p^6 \ln(p)^2 + \frac{608}{59049}p^{2/3} \ln(p) \nonumber\\ && + \frac{3584}{6561}p^{10/3} \ln(p)+ \frac{128}{6561}p^{10/3} \ln(p)^2+ \frac{64}{59049}p^{2/3} \ln(p)^2
\nonumber\\ && + \frac{436}{19683}p^{2/3} -\frac{32}{243} \ln(p) p^6 , \nonumber\\
g_2(p) &=& \frac{64}{81}p^9\ln(p)^2 + \frac{436}{531441}p+ \frac{14036}{531441} p^{11/3} +\frac{2164}{19683}p^{19/3}
\nonumber\\ &&-\frac{100}{729}p^9 -\frac{32}{243}p^9 \ln(p)+ \frac{320}{6561}p^{19/3}\ln(p)^2 +\frac{1184}{2187}p^{19/3} \ln(p)
\nonumber\\ && +\frac{5408}{177147}p^{11/3} \ln(p) +\frac{320}{177147}p^{11/3} \ln(p)^2 +\frac{608}{1494323}p \ln(p)\nonumber\\ && +\frac{64}{1594323}p \ln(p)^2,   \nonumber\\
g_3(p) &=& \frac{64}{81}p^{12}\ln(p)^2
+\frac{640}{177147}p^{20/3}\ln(p)^2 +\frac{688}{6561}p^{28/3}
+\frac{7232}{4782969}p^4 \ln(p) \nonumber\\ &&
+\frac{512}{6561}p^{28/3} \ln(p)^2+ \frac{25808}{14348907}p^4
+\frac{3520}{6561}p^{28/3} \ln(p) +\frac{200}{6561}p^{20/3}
\nonumber\\ && +\frac{512}{4782969}p^4 \ln(p)^2
+\frac{436}{14348907}p^{4/3} +\frac{608}{43046721}p^{4/3}\ln(p)
\nonumber\\ && +\frac{64}{43046721}p^{4/3}\ln(p)^2
-\frac{32}{243}p^{12}\ln(p) +\frac{8960}{177147}p^{20/3}\ln(p)
-\frac{100}{729}p^{12}.
\end{eqnarray}

We have plotted DM, for a single particle Abe configuration in Fig. (\ref{fig13}) and scalar curvature in
Fig. (\ref{fig14}).  In this case, the system remains well defined and stable, except near $q=1$, where there are large metric fluctuations, as seen in Fig. (\ref{fig13}).
On the other hand,  the system becomes more stable with increasing  number of particles, as can be seen from the Figs. (\ref{fig15}) and (\ref{fig17})), where the local correlations,
as depicted by the metric tensor, can be seen to display less fluctuations (bumps in the curve) when compared to the single particle case of Fig. (\ref{fig13}).
The system is regular except near $q\sim 0.1$,  as seen in  Fig. (\ref{fig18}).

\subsection{Structural Entropy}

The structural entropy has been defined as \cite{kz03}
\begin{eqnarray}
S_s &=& S - (S_2^R) \nonumber\\
&=& -\sum\limits_i p_i \ln p_i + \ln(\sum\limits_i p_i^2),
\label{structural}
\end{eqnarray}
where $S$  is the standard Shannon entropy  (\ref{shannon}) while
$S_2^R$ is the R\'{e}nyi entropy (\ref{renyi})  for $q=2$. This
entropy was studied in \cite{pv92}, where it was  shown that an
increase in the structural entropy,  in the  case  of  a tight
binding  model,  indicates  Anderson localization.

Components of the thermodynamic metric, in the case of  the
structure entropy for an arbitrary number of particles (say $n$)
are as follows,
\begin{eqnarray}
g_{p_i,p_j}&=&\frac{4 p_i p_j}{(\sum_{i} p^2_i)^2}, \nonumber\\
g_{p_i,p_i}&=& \frac{1}{p_i}-\frac{2}{\sum_i p_i^2}+ \frac{4
p_i^2}{(\sum_i p_i^2)^2},
\end{eqnarray}
where $i, j$ will take values from $1$-$n$. In this article we
shall demonstrate the three cases, {\it viz.}, two particle, three
particle and four particle configurations.

\subsubsection{Two particle configuration}
The determinant of the thermodynamic metric is
\begin{equation}
\Vert g \Vert = \frac{p_1^4+p_2^4+2(p_1^3+p_2^3)+2 p_1^2 p_2^2-2(p_2 p_1^2+p_2^2 p_1)-4p_1p_2 }{p_1 p_2 (p_1^2 +p_2^2)^2},
\end{equation}
while the scalar curvature is
\begin{equation}
R=-2 \bigg(\frac{p_1 p_2(-2p_2 p_1^2-2p_2^3+p_1^4+2p_1^2p_2^2-2p_1
p_2^2-2p_1^3 +p_2^4)} {(p_1^4+p_2^4+2(p_1^3+p_2^3)+2p_1^2
p_2^2-2(p_2 p_1^2+p_2^2 p_1)-4p_1p_2)^2}\bigg).
\end{equation}

\begin{figure}
\includegraphics[width=7.0cm]{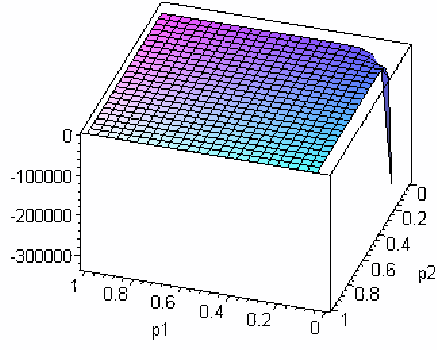}
\caption{Determinant of the metric tensor as a function of the probabilities $p_1$ and $p_2$, in a two particle structural configuration.} \label{fig19}
\vspace*{0.5cm}
\end{figure}

\begin{figure}
\includegraphics[width=7.0cm]{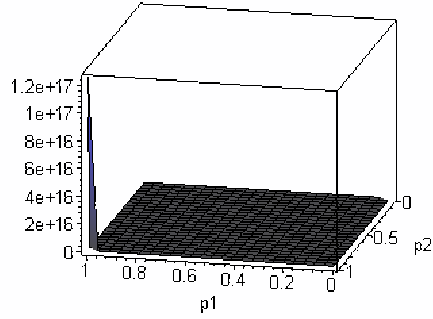}
\caption{Thermodynamic curvature as a function of the probabilities $p_1$ and $p_2$, in a two particle structural configuration.} \label{fig20}
\vspace*{0.5cm}
\end{figure}

\subsubsection{Multi-particle configuration}
For $p_i=p$ ($i=1$-$3$), the determinant of the metric tensor and the scalar curvature are,
\begin{equation}
\Vert g \Vert =\frac{1}{27p^6} \bigg(27 p^3-18 p^2- 12 p+8\bigg),
\end{equation}
and
\begin{equation}
 R=2 \bigg(\frac{81 p^{4}+216 p^3-396 p^{2}+192 p-32}{(27 p^3-18 p^2- 12 p+8)^2}\bigg),
\end{equation}
respectively.
In the four particle  case ($i=1$-$4$), the determinant of the metric tensor is
\begin{equation}
\Vert g \Vert =\frac{1}{16p^8} \bigg(16 p^4-16p^3+4 p^2-1\bigg),
\end{equation}
while the scalar curvature is
\begin{equation}
 R=-\frac{1}{128} \bigg(\frac{
384-3840 p+15360 p^{2}-27648 p^{3}+12288 p^{4}+24 576 p^{5}-24576
p^{6}}{(16 p^4-16p^3+4 p-1)^2}\bigg).
\end{equation}

\begin{figure}
\includegraphics[width=7.0cm]{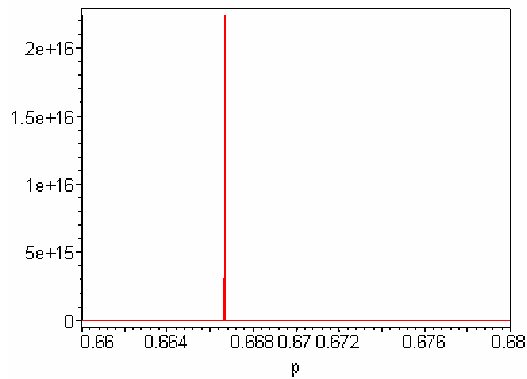}
\caption{Thermodynamic scalar curvature as a function of probability  $p = p_i$, for all $i$, in a three particle structural configuration.} \label{fig21}
\vspace*{0.5cm}
\end{figure}
The determinant of the metric tensor (DM) and scalar curvature, for a two particle structural configuration, are depicted in
Figs. (\ref{fig19}) and (\ref{fig20}), as a function of the probabilities $p_1$ and $p_2$. From Fig.  (\ref{fig19}), it is easy to see that the system
becomes locally unstable and ill-defined near
($p_1=0$, $p_2=0$).  From Fig. (\ref{fig20}), the  system is seen to be globally stable and regular, except ($p_1=1$, $p_2=1$). The multi-particle systems
show a similar nature. The scalar curvature is well-defined and regular, except for a
singularity arising from the zeros of the metric tensor,  as seen in  Fig. (\ref{fig21}) for $p_i = 0.6675$.

\section{Geometric interpretation of the additivity of entropies}

An interesting observation, following from the  geometric analysis presented, is that the additivity of R\'{e}nyi  and pseudo-additivity
of Tsallis entropies may be analyzed, in a simple way,  by considering
\begin{eqnarray}
S^{R}_{q,I,II} - (S^{R}_{q,I}  + S^{R}_{q,II} ) &=& 0,\nonumber\\
S^{R}_{q,I,II} - (S^{T}_{q,I}  + S^{T}_{q,II} ) &=&  -(q - 1) S^{T}_{q,I}   S^{T}_{q,II},
\end{eqnarray}
where, R denotes  R\'{e}nyi and T denotes  Tsallis entropies,
respectively. I and II denote the subsystems and q is the entropic
parameter,  as in the earlier sections. The covariant
characterization, which the present analysis explores, would bring
out  the geometric meaning of additivity for R\'{e}nyi  and
pseudo-additivity for Tsallis systems. More precisely, we can
covariantly accomplish these statistical properties simply by
considering the difference of local and global correlations, that
is the metric tensor and thermodynamic scalar curvature,
respectively.

Corresponding to the determinant of the metric tensor and the scalar curvature equations,
for the R\'{e}nyi and Tsallis configurations, we may realize the additivity and the pseudo-additivity at a given temperature for fixed number of probabilities.
In order to have a  global characterization, we
define $\Delta R =R^{Renyi} - R^{Tsallis} $. It is worth mentioning that $\Delta R$ involves both the concept of additivity of the R\'{e}nyi configurations and pseudo-additivity of the Tsallis configurations. Hence, for a given set of configuration parameters $ (q, p_i) $, the plot of the quantity $\Delta R $,  as in Fig. (\ref{fig23}),  brings out the
order of  non-additivity,
at a global level of the correlation volume of the concerned statistical system. It may further be
envisaged that the quantity $ \Delta R $ also indicates potential phase transitions, if any,  in  the chosen, finite parameter, system modulating over
a range of temperatures.

\begin{figure}
\includegraphics[width=9.0cm]{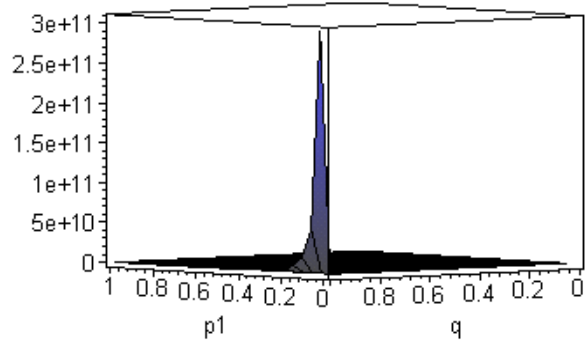}
\caption{$\Delta \Vert g \Vert$ as a function of $p_1$ and q.} \label{fig22}
\vspace*{0.5cm}
\end{figure}

\begin{figure}
\includegraphics[width=9.0cm]{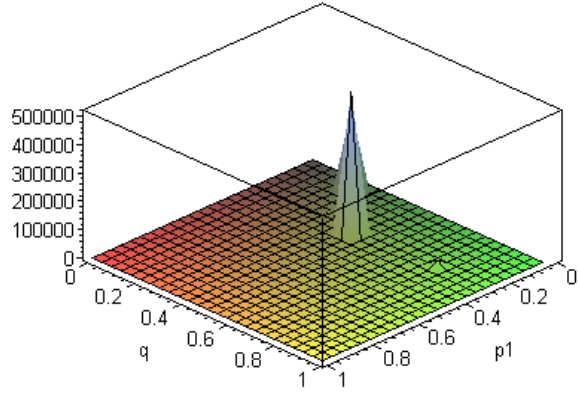}
\caption{$\Delta R$ as a function of $p_1$ and q} \label{fig23}
\vspace*{0.5cm}
\end{figure}

On the other hand,  the difference $\Delta \Vert g \Vert = \Vert g \Vert^{Renyi} -  \Vert g \Vert^{Tsallis}$,  as shown in
Fig. (\ref{fig22}),  would be a measure of the stability,  with respect to
non-additivity, of fluctuations in the configuration  over the parameters under consideration.
We observe that in the  domain of parameters, the intrinsic geometric notion of non-additivity may locally be analyzed via the statistical correlations defined over an
equilibrium distribution. In particular,  the component equations $\Delta g_{ij} =g^{Renyi}_{ij} - g^{Tsallis}_{ij}$ define an order of non-additivity at
an intrinsic metric level. This local information would further put forward the picture of non-additivity,  as a result of local statistical fluctuations over an
equilibrium thermodynamic characterization.  Thus, from the present investigations, involving
the calculation of the metric tensor and  scalar
curvatures we find that, for given entropic
parameter $q$ and probabilities $ p_i $,   the associated
differences  $\Delta \Vert g \Vert$ and $ \Delta R$,  graphically depicted in
Figs. (\ref{fig22}),  (\ref{fig23}),  respectively, bring
out the nature of  statistical fluctuations at the local as well as global level.

\section{Conclusion and Future prospects}

In this paper, we have applied thermodynamic geometry to open
statistical systems. The metric tensor, $ g_{ij} $, is an
indicator of local correlations in the system and can be used to
explore questions related to its local thermodynamic stability,
while a non-zero thermodynamic scalar curvature, $R$,  is a
signature of global correlations in the system and would be useful
to address questions related to phase transitions in it.
Specifically, a zero scalar curvature would indicate that the
components of the statistical system fluctuate independent of each
other, while a divergent scalar curvature would be an indicator of
phase transitions. We have analyzed the, well known,
Gibbs-Shannon, R\'{e}nyi and Tsallis entropies. From our
motivation to understand complex statistical models, we also study
the intrinsic thermodynamic geometry of Abe and Structural
entropy.

Similar conclusions are valid, at a qualitative level, for the two and three
parameter R\'{e}nyi configurations, and is depicted in Figs.  (\ref{fig3})-(\ref{fig6}).
The negative scalar curvature is an indicator of the attractive nature of the system, as indicated by the
crusts of Fig. (\ref{fig4}), for the two particle case. Following Fig. (\ref{fig6}), we observe that the three parameter
system has larger attraction, as it possesses  larger negative valued
scalar curvature.

In order to make detailed predictions, we have made an analysis for  one, two
and three parameter, thermally excited configurations. We find that the local
statistical correlation functions associated with the Gibbs-Shannon, R\'{e}nyi,
Tsallis, Abe and the structural configurations are well defined, and diagonal
components correspond to definite heat capacity expressions. We observe
 that the nature of correlations remain the same, qualitatively, as we increase
the number of parameters in either configuration.

For  few parameter systems, we find that the intrinsic
thermodynamic system has a non-zero scalar curvature. Physically,
this shows that the generic R\'{e}nyi, Tsallis, Abe and structural
configurations correspond to an interacting statistical system. On
the other hand, the Gibbs-Shannon entropy, having a zero scalar
curvature, corresponds to a non-interacting statistical system.
Thus, we are able to give a geometric meaning to the various
entropies.


Further, the case of  power law Hamiltonians and their quantum
mechanical counterparts arise interestingly. The paper concludes
with the following perspective study.

%
\subsection{Power Law Hamiltonians}
To extend the geometric approach, we may start with the power law Hamiltonian with
its dependence on a parameter $x$ as
\begin{equation}
H_i = C x_i^{\chi}.   \label{power}
\end{equation}
Such Hamiltonians are useful in modeling the thermostatistics of  complex systems \cite{bash}.
When $\chi  = 2$, this configuration corresponds to the one-dimensional ideal gas with quadratic
hamiltonian $H = p^2/2m$. For such models \cite{bash}, the R\'{e}nyi distribution is given by
\begin{equation}
p_i^R = Z^{-1} \Big(1 - \frac{q-1}{\chi q}
(C_u x_i^k - 1)\Big)^{\frac{1}{q-1}}.   \label{rendispower}
\end{equation}
Here $C_u = C/U$. where $C$ is defined as in (Eqn. \ref{power})
and $U$ is the average energy. Substituting Eq.
(\ref{rendispower}) into the Eq. (\ref{RenyiEntropy}), the
R\'{e}nyi entropy can be obtained for $q
> q_{min} = 1/(1+\chi)$.

For $\chi= 2$, we have extensively analyzed the thermodynamic
geometric properties for these systems in Section III. In fact,
the generic nature of local and global correlations, arising from
the determinant of the metric tensor and scalar curvature, as
depicted in the respective plots, possess a very similar nature
under an addition of the extra variable $p_i$. It is worth
mentioning, that we find similar issues to hold in various
possible cases of the open statistical configurations with $\chi
= 2$.

The R\'{e}nyi and Shannon entropies \cite{bash} have been
investigated further,  for similar considerations, for $q >
q_{min}$. The corresponding Tsallis, Abe and structural entropies,
are expected to show similar thermodynamic geometric behavior.
This analysis brings out that the generic open systems have
well-defined, interacting statistical configurations.

For a given statistical system with a specific entropic parameter $q$, the determinants
of the metric tensor and the corresponding scalar curvatures are envisaged to
have definite connections with the superstatistics \cite{bash, cb01, bc03, bash2},
where the entropic parameter is defined by physical properties of a complex system,  that can exchange
energy and heat with a thermostat. These issues would be the subject  of future investigations.

\subsection{Quantum mechanical counterpart}

As previously mentioned, the treatment followed in this paper is
classical. It would be pertinent to develop the corresponding
geometric treatment in the quantum  regime.  From this
perspective \cite{Provost}, one may explore issues concerning
entanglement, long-range global correlations and their near
equilibrium behavior, from the perspective of thermodynamic
geometry.

\vspace*{1.0cm}

{\bf Acknowledgement:} BNT thanks the Prof. V. Ravishankar, Prof.
S. Bellucci, Prof. P. Jain, Prof. U. B. Tewari, Prof. M. K.
Harbola, Prof. R. K. Thareja and Prof. S. G. Dhande for their
support and encouragement while this work was going on. VC and BNT
would further like to acknowledge CSIR New Delhi, India for
financial support.

\end{document}